\documentclass[10pt]{article}

\usepackage{amstext,amsgen,latexsym,amsmath}
\usepackage{amstext,amssymb,amsfonts}
\usepackage{theorem}
\usepackage{pifont}

 \newcommand{\bs}{\bigskip}
 \newcommand{\ms}{\medskip}
 \newcommand{\n}{\noindent}
 
 \newcommand{\hs}[1]{\hspace*{ #1 mm}}
 \newcommand{\vs}[1]{\vspace*{ #1 mm}}

 \newcommand{\setempty}{{\mathrm{\O}}}
 \newcommand{\real}{{\mathbb{R}}}

 \newcommand{\nat}{{\mathbb{N}}}
 \newcommand{\dyadic}{{\mathbb{D}}}
 \newcommand{\integers}{{\mathbb{Z}}}
 \newcommand{\rational}{{\mathbb{Q}}}
 
 \newcommand{\complex}{{\mathbb{C}}}
 \newcommand{\ptcomplex}{{\tilde{\mathbb{C}}}}
 
 \newcommand{\algebraic}{{\mathbb{A}}}

 \newcommand{\vsigma}{\mbox{\boldmath $\sigma$}}
 \newcommand{\vtau}{\mbox{\boldmath $\tau$}}
 
 \newcommand{\vepsilon}{\mbox{\boldmath $\epsilon$}}
 \newcommand{\vd}{\mbox{\boldmath $d$}}
 \newcommand{\svsigma}{\mbox{\boldmath ${}_{\sigma}$}}
 \newcommand{\svtau}{\mbox{\boldmath ${}_{\tau}$}}
 
 \newcommand{\svepsilon}{\mbox{\boldmath ${}_{\epsilon}$}}
 \newcommand{\svd}{\mbox{\boldmath ${}_{d}$}}
 \newcommand{\svk}{\mbox{\boldmath ${}_{k}$}}

 \newcommand{\vk}{\mbox{\boldmath $k$}}
 \newcommand{\vx}{\mbox{\boldmath $x$}}

 \newcommand{\dom}{{\mbox{dom}}}
 
 \newcommand{\ran}{{\mbox{ran}}}

 \newcommand{\co}{\mathrm{co}\mbox{-}}

 \newcommand{\ie}{\textrm{i.e.},\hspace*{2mm}}
 \newcommand{\eg}{\textrm{e.g.},\hspace*{2mm}}
 
 \newcommand{\etalc}{\textrm{et al.}}

 \newcommand{\CC}{{\cal C}}
 \newcommand{\FF}{{\cal F}}

 \newcommand{\GG}{{\cal G}}
 
 \newcommand{\PP}{{\cal P}}
 \newcommand{\QQ}{{\cal Q}}

 \newcommand{\p}{{\mathrm{P}}}
 \newcommand{\np}{{\mathrm{NP}}}
 
 \newcommand{\rp}{{\mathrm{RP}}}
 
 \newcommand{\bpp}{{\mathrm{BPP}}}
 
 \newcommand{\pp}{{\mathrm{PP}}}
 \newcommand{\up}{{\mathrm{UP}}}
 
 \newcommand{\pspace}{{\mathrm{PSPACE}}}
 \newcommand{\e}{{\mathrm{E}}}

 \newcommand{\cequalp}{{\mathrm{C}}_{=}{\mathrm{P}}}
 \newcommand{\spp}{{\mathrm{SPP}}}

 \newcommand{\eqp}{{\mathrm{EQP}}}
 \newcommand{\nqp}{{\mathrm{NQP}}}
 
 \newcommand{\bqp}{{\mathrm{BQP}}}
 \newcommand{\zqp}{{\mathrm{ZQP}}}
 \newcommand{\pqp}{{\mathrm{PQP}}}
 \newcommand{\wqp}{{\mathrm{WQP}}}
 \newcommand{\qma}{{\mathrm{QMA}}}

 \newcommand{\fp}{{\mathrm{FP}}}
 \newcommand{\npsv}{{\mathrm{NPSV}}}
 
 \newcommand{\sharpp}{\#{\mathrm{P}}}
 \newcommand{\sharpe}{\#\mathrm{{E}}}
 \newcommand{\optp}{{\mathrm{OptP}}}
 \newcommand{\spanp}{{\mathrm{SpanP}}}
 \newcommand{\gapp}{{\mathrm{GapP}}}

 \newcommand{\feqp}{{\mathrm{FEQP}}}
 \newcommand{\fbqp}{{\mathrm{FBQP}}}
 \newcommand{\qmasv}{{\mathrm{QMASV}}}
 \newcommand{\sharpqp}{\#{\mathrm{QP}}}
 \newcommand{\gapqp}{{\mathrm{GapQP}}}

 \newcommand{\appin}{\underset{\widetilde{\hs{3}}}{\in}^{p}}
 
 \newcommand{\appsubseteq}{\Subset^{p}}
 \newcommand{\stappsubseteq}{\Subset^{e}}

 \newcommand{\low}{{\mathrm{low\mbox{-}}}}

\theoremstyle{plain}
\theoremheaderfont{\bfseries}
 \newtheorem{theorem}{Theorem}[section]
 \newtheorem{lemma}[theorem]{Lemma}
 \newtheorem{proposition}[theorem]{Proposition}
 \newtheorem{corollary}[theorem]{Corollary}

 {\theorembodyfont{\rmfamily}
       \newtheorem{definition}[theorem]{Definition}}
 {\theorembodyfont{\rmfamily} }
 {\theorembodyfont{\rmfamily} }

 {\theorembodyfont{\rmfamily}
       }
 {\theorembodyfont{\rmfamily}
       }

 \newenvironment{proof}{\par \noindent
            {\bf Proof. \hs{2}}}{\hfill$\Box$ \vspace*{2mm}}

 \newenvironment{proofof}[1]{\par \noindent
         {\bf Proof of #1.\hs{2}}}{\hfill$\Box$ \vspace*{2mm}}

 \newcommand{\ceilings}[1]{\lceil #1 \rceil}
 \newcommand{\floors}[1]{\lfloor #1 \rfloor}
 \newcommand{\pair}[1]{\langle #1 \rangle}

 \newcommand{\qubit}[1]{| #1 \rangle}
 \newcommand{\bra}[1]{\langle #1 |}
 \newcommand{\ket}[1]{| #1 \rangle}
 \newcommand{\measure}[2]{\langle #1 | #2 \rangle}

\begin{document}

\begin{center}
{\Large {\bf Analysis of Quantum Functions\footnote{A preliminary 
version appeared in the Proceedings of the 19th International
Conference on Foundations of Software Technology and Theoretical
Computer Science, Lecture Notes in Computer Science, Springer-Verlag,
Vol.1738, pp.407--419, 1999.}}} \bs\\
{\sc Tomoyuki Yamakami}\footnote{Most of this work was done
while the author was at the Department of Computer Science in Princeton
University between September 1997 and June 1999 and was partly
supported by NSERC Fellowship as well as DIMACS Fellowship.} \ms\\
{School of Information Technology and Engineering\\ 
University of Ottawa, Ottawa, Ontario, Canada K1N 6N5} 
\end{center}
\bs

\paragraph{Abstract:}
This paper initiates a systematic study of quantum functions, which
are (partial) functions defined in terms of quantum mechanical
computations. Of all quantum functions, we focus on resource-bounded
quantum functions whose inputs are classical bit strings. We prove
complexity-theoretical properties and unique characteristics of these
quantum functions by recent techniques developed for the analysis of
quantum computations. We also discuss relativized quantum functions
that make adaptive and nonadaptive oracle queries.

\ms
\n{\sf key words:} quantum function, 
quantum Turing machine, nonadaptive query, 
oracle separation

\section{Overture}

A paradigm of a quantum mechanical computer was first proposed in the
1980s \cite{Beni80,Deu85,Fey82} to exercise more computational power
over the silicon-based computer, whose development is speculated to
face a physical barrier. Since quantum mechanics is thought to govern
Nature, a computer built upon quantum physics is of great importance.
A series of discoveries of fast quantum algorithms in the 1990s
\cite{Gro96,Sho97} has raised enthusiasm among computer scientists as
well as physicists. These discoveries have since then supplied general
and useful tools in programming quantum algorithms.

A quantum computer has been mathematically modeled in several
different manners, including quantum Turing machines
\cite{BV97,Deu85}, quantum circuits \cite{Deu89,Yao93}, and
topological computations \cite{FKLW01}. This paper uses a multiple
tape model of {\em quantum Turing machine} (referred to as {\em QTM})
along the line of expositions \cite{ADH97,BV97,NO02,ON99,Yam99,YY99}
due to its close connection to a classical off-line Turing machine
(TM, for short). A quantum computation of a QTM is a series of
superpositions of the machine's configurations whose evolution obeys
quantum physics. An evolution of such a superposition allows any
computation path to interfere with other computation paths. This
phenomenon is known as {\em quantum interference} and a QTM can
exploit quantum interference to achieve a large volume of parallel
computations efficiently. Such a machine naturally computes a
(partial) function.  For instance, the Integer Factorization 
Problem (\ie given a positive integer, find its factors)
is solved by Shor's polynomial-time quantum algorithm \cite{Sho97}. 
Any function that can be defined in terms of quantum mechanical 
computations is in genarl referred to as a {\em quantum function}. 
 
A study of function classes has been an important subject in classical
complexity theory. Search problems and optimization problems are in
fact functions and are of special interest in many practical areas of
computer science. The treatment of functions is,
however, slightly different from that of languages (or simply called
{\em sets}) because of the size of output bits. 
Since the 1960s, researchers have investigated
various function classes, including $\fp$, $\sharpp$ \cite{Val79},
$\npsv$ \cite{BLS84,SMB83}, $\optp$ \cite{Kre88}, $\spanp$
\cite{KST89}, and $\gapp$ \cite{FFK94}. Similarly, we need to
develop a general theory of quantum functions. Of all
quantum functions, this paper focuses only on those whose inputs are
classical bit strings since focusing on classical inputs makes it
possible for us to relate quantum functions to classically computable
functions in numerous ways. Our goal is thus to establish the
foundations of the theory of quantum functions by conducting a
systematic study of the behaviors of quantum functions and by
exploring the similarities and differences between classical functions
and quantum functions.

We consider two categories of quantum functions. A {\em quantum
computable function} computes an output of a QTM with high
probability. Such functions have been used in the literature without
proper names.  Let $\feqp$ and $\fbqp$ denote respectively
the collections of all
functions computed in polynomial time 
by certain well-formed QTMs with
certainty and with probability at least $3/4$.
These function classes are viewed as quantum generalizations of the
class of polynomial-time classically computable functions. Similarly,
a partial single-valued $\qma$-function is a quantum variant of a
single-valued $\np$-function. A {\em quantum probability function}, in
contrast, computes the acceptance probability of a well-formed
QTM. For notational convenience, $\sharpqp$ ({\em sharp QP}) denotes
the collection of such quantum functions particularly witnessed by
polynomial-time well-formed QTMs. An important variant of such a
function is the one that computes the gap between the acceptance and
rejection probabilities of a well-formed QTM. 
We call such functions {\em quantum
probability gap functions} and use the notation $\gapqp$ to denote the
collection of all polynomial-time quantum probability gap
functions. We show that $\gapqp$ is the subtraction closure of
$\sharpqp$.

There have been developed several proof techniques in quantum
complexity theory during the 1990s. These techniques are crucial to
our analysis of quantum functions. An amplitude amplification
technique of Brassard, H{\o}yer, and Tapp \cite{BHT98}, for instance,
is used to show that any $\sharpqp$-function can be closely
approximated by a certain $\fbqp$-function.  Refining an idea of
Fenner, Green, Homer, and Pruim \cite{FGHP99}, we show a striking
feature of quantum probability gap functions: if $f\in\gapqp$ then
$f^2\in\sharpqp$. Based on a series of results by Adleman, DeMarrais,
and Huang \cite{ADH97} and Yamakami and Yao \cite{YY99}, we draw the
close connection between $\gapqp$-functions and $\gapp$-functions. In
particular, if all amplitudes are restricted to algebraic numbers, the
sign (\ie positive, zero, or negative) of the value of a
$\gapqp$-function is shown to coincide with that of a certain
$\gapp$-function.  This relationship further brings a new
characterization of $\pp$ in terms of $\gapqp$-functions. As an
immediate consequence, the quantum analogue of $\pp$ called $\pqp$
with algebraic amplitudes collapses to $\pp$.

To enhance a computation of a QTM, we further allow the machine to
access an oracle by way of oracle queries. An oracle quantum
computation dates back to Deutsch and Jozsa \cite{DJ92}, who showed
that a quantum query can receive more information from an oracle than
a classical query does.  Generally, one oracle query depends on its
previous oracle answers. This pattern of oracle accesses is
categorized as {\em adaptive queries}. On the contrary, {\em
nonadaptive queries}\footnote{Our quantum nonadaptive query model
seems different from a quantum analogue of a truth-table reduction,
which is widely used as a nonadaptive query model in a classical
setting.} (or parallel queries) refer to the case where an oracle QTM
prepares a list of query words along each computation path before
making the first query in the entire computation. For a nonadaptive
query case, we use the notation $\feqp_{\|}^{A}$ to denote the
collection of all $\feqp^A$-functions that make nonadaptive queries to
oracle $A$.

In a classical bounded query model, a function class and a language
class generally behave in different manners; for instance,
$\fp_{\|}^{\np}$ is believed to differ from $\fp^{\np[O(\log n)]}$
whereas $\p_{\|}^{\np}$ coincides with $\p^{\np[O(\log n)]}$
\cite{Wag90}. In contrast, quantum interference makes it possible to
draw such functions and languages close together by a use of the
quantum algorithm of Bernstein and Vazirani \cite{BV97}. Moreover, we
exhibit an oracle that separates $\feqp_{\|}^A$ from $\fp^A$ and also
construct another oracle $B$ that makes $\fp^B$ harder than
$\feqp_{\|}^B$.  These relativized results together imply that
$\eqp_{\|}^{A}\nsubseteq\p^A$ and $\p^{B}\nsubseteq\eqp_{\|}^B$. This
exemplifies a peculiar nature of quantum nonadaptive queries.

The study of quantum functions finds useful applications to decision
problems. We show a relationship between the $\eqp=?\pp$ question and
the closure property of $\sharpqp$ under the maximum and minimum
operators. In the course of our study, we introduce the new quantum
complexity class $\wqp$ ({\em wide QP}), which naturally expands $\up$
and $\eqp$. An oracle of Fortnow and Rogers \cite{FR99} can separate
$\wqp$ from $\eqp$.

Our investigation merely opens a door to a largely uncultivated area
of quantum functions in quantum complexity theory. As our study
unfolds, we nevertheless leave unanswered more questions on the
behaviors of quantum functions. We strongly hope that a vigorous study
of quantum functions will bring us the answers to these questions in
the future.

\section{Basic Notions and Notation}\label{sec:notions}

We briefly introduce fundamental notions and notation necessary to
read through this paper.

Denote by $\nat$ and $\integers$, respectively, the set of all natural
numbers (that is, non-negative integers) and the set of all
integers. Set $\nat^{+}=\nat-\{0\}$. For each $d\in\nat^{+}$, let
$\integers_{d}=\{0,1,\ldots,d-1\}$ and
$\integers_{[d]}=\{-d,\ldots,-1,0,1,\ldots,d\}$. Moreover, let
$\rational$, $\real$, and $\complex$ be the sets of all rational
numbers, real numbers, and complex numbers, respectively. In this
paper, the notation $\algebraic$ is used to denote the set of {\em
complex algebraic numbers}.
 
The notation $[a,b]$ denotes the real interval between $a$ and
$b$. Similarly, we use $(a,b]$ and $(a,b)$. For any finite set $S$,
$|S|$ denotes the {\em cardinality} of $S$. We say that, for any
infinite set $S$, a property $\PP(x)$ {\em holds for almost all $x$ in
$S$} if $\{x\in S\mid \mbox{$\PP(x)$ does not hold}\}$ is a finite
set.

A finite set $\Gamma=\{\gamma_1,\gamma_2,\ldots,\gamma_k\}$ of complex
numbers is said to be {\em linearly independent} if
$\sum_{i=1}^{k}a_i\gamma_i\neq0$ for any non-zero $k$-tuple
$(a_1,a_2,\ldots,a_k)\in \rational^k$ and $\Gamma$ is {\em
algebraically independent} if
$q(\gamma_1,\gamma_2,\ldots,\gamma_k)\neq0$ for any function
$q\in\rational[x_1,x_2,\ldots,x_k]$ that is not identically $0$. For
any subset $A$ of $\complex$, $\rational(A)$ denotes the field
generated by all elements in $A$ over $\rational$. In this paper, a
{\em polynomial with $k$ variables} means an element in
$\nat[x_1,x_2,\ldots,x_k]$ and thus, all polynomials are assumed to be
nondecreasing.

We often use the $\lambda$-notation to describe functions. The
notation $\lambda x.f(x)$ means the function $f$ itself. For example,
$\lambda x.(2x+3)$ denotes the function that outputs $2x+3$ on input
$x$.  For any (partial) function $f$, $\dom(f)$ and $\ran(f)$ denote 
respectively the {\em domain} and the {\em range} of $f$. We write
$f(x)\!\uparrow$ to mean that $f(x)$ is {\em undefined} (\ie
$x\not\in\dom(f)$) and we also write $f(x)\!\downarrow$ if $f(x)$ is
{\em defined} (\ie $x\in\dom(f)$). For any class $\FF$ of partial
functions, the {\em domain} of $\FF$ is $\dom(\FF)=\{\dom(f)\mid
f\in\FF\}$. For any two functions $f$ and $g$ with the same domain,
$f-g$ denotes $\lambda x.(f(x)-g(x))$.

For simplicity, we use a binary alphabet $\Sigma=\{0,1\}$ throughout
this paper unless otherwise stated. For any string $x$, the {\em
length} of $x$, denoted $|x|$, is the number of bits in $x$.  For any
number $n\in\nat$, $\Sigma^n$ ($\Sigma^{\leq n}$, $\Sigma^{\geq n}$,
resp.) represents the set of all strings of length $n$ ($\leq n$,
$\geq n$, resp.). Let $\Sigma^*=\bigcup_{n\in\nat}\Sigma^n$. A subset
of $\Sigma^*$ is called a {\em language} or simply a {\em set}. Any
collection of certain languages or functions is conventionally called
a {\em complexity class}. For any subsets $A$ and $B$ of $\Sigma^*$,
$\overline{A}$ denotes $\Sigma^*-A$ (the {\em complement} of $A$),
$A\oplus B$ is $\{0x\mid x\in A\}\cup\{1x\mid x\in B\}$ (the {\em
disjoint union} of $A$ and $B$), and $A\triangle B$ is $(A-B)\cup
(B-A)$ (the {\em symmetric difference} of $A$ and $B$). For any
language class $\CC$, $\co\CC$ denotes the class of the complements of
any sets in $\CC$.  For any set $S$, its {\em characteristic function}
$\chi_{S}$ is defined as $\chi_{S}(x)=1$ if $x\in S$ and
$\chi_{S}(x)=0$ otherwise.

Let $\nat^{\Sigma^*}$ be the set of all functions that map $\Sigma^*$
to $\nat$. Similarly, we define $\nat^{\nat}$, $\{0,1\}^{\Sigma^*}$,
$[0,1]^{\Sigma^*}$, etc.  A function $f$ from $\Sigma^*$ to $\Sigma^*$
($\nat$, resp.) is {\em polynomially bounded} if there exists a
polynomial $p$ such that $|f(x)|\leq p(|x|)$ ($f(x)\leq p(|x|)$,
resp.) for all $x$ in $\Sigma^*$. A function $f$ from $\Sigma^*$ to
$\Sigma^*$ is {\em length-regular} if, for every pair
$x,y\in\Sigma^*$, $|x|=|y|$ implies $|f(x)|=|f(y)|$. For any two
functions $f,g\in [0,1]^{\Sigma^*}$ and any function $\epsilon\in
[0,1]^{\nat}$, we say that $f$ {\em $\epsilon(n)$-approximates} $g$ if
$|f(x)-g(x)|\leq \epsilon(|x|)$ for almost all $x$ in $\Sigma^*$.  Let
$\FF$ and $\GG$ be any subsets of $[0,1]^{\Sigma^*}$.  For any
function $f\in [0,1]^{\Sigma^*}$, we write $f\appin\FF$ if, for every
polynomial $p$, there exists a function $g\in\FF$ that
$1/p(n)$-approximates $f$. The notation $\FF\appsubseteq\GG$ means
that $f\appin\GG$ for all functions $f$ in $\FF$. Similarly, the
notation $\FF\stappsubseteq\GG$ is defined using
``$2^{-p(n)}$-approximation'' instead of ``$1/p(n)$-approximation.''

We freely identify any natural number with its binary representation
throughout this paper.  When we discuss integers, we also identify
each integer with its binary representation following a {\em sign bit}
that indicates the (positive or negative) sign\footnote{For example,
we set 1 for a positive integer and 0 for a negative integer. For
uniqueness, the integer $0$ always has a positive sign.} of the
integer. An integer with such a representation is called a {\em binary
integer} for convenience.  A rational number is also identified as a
pair of integers, which are further identified as binary integers.

As a mathematical model of classical computation, we use a
multiple-tape off-line TM with two-way infinite read/write tapes whose
cells are indexed by $\integers$. A cell indexed $0$, on which all
tape heads rest at the start of a computation, is called the {\em
start cell}. We use deterministic, nondeterministic, and probabilistic
TMs.  In addition, a {\em reversible TM} is a deterministic TM for
which each configuration has at most one predecessor
\cite{Ben73,BV97}.  All TMs can move its heads to the right and to the
left and also allow them to stay still. For any nondeterministic TM
$M$ and any string $x$, the notation $\#M(x)$ ($\#\overline{M}(x)$,
resp.) denotes the total number of accepting (rejecting, resp.)
computation paths of $M$ on input $x$.
 
The following lemma is useful in order to simulate a classical
computation on a QTM.

\begin{lemma}\label{lemma:reversible}{\rm \cite{Ben73,BV97}}\hs{2}
Any deterministic TM $M$ that, on any input $x$, outputs a string
$M(x)$ on an output can be simulated with polynomial slowdown by a
certain reversible TM $N$ such that (i) on any input $x$, $N$ outputs
$(x,M(x))$ onto an output tape, (ii) all the heads of $N$ move back to
their start cells, and (iii) the running time of $N$ on input $x$
depends only on the lengths of both input $x$ and output $M(x)$.
\end{lemma}

Let $\ptcomplex$ denote the set of all {\em polynomial-time
approximable complex numbers}, \ie complex numbers whose real and
imaginary parts are deterministically approximated to within $2^{-n}$
in time polynomial in $n$.  A {\em dyadic rational number} is the
number of the form $x.y$ for certain finite binary strings $x$ and
$y$, and $\dyadic$ denotes the set of all dyadic rational
numbers. Note that $\nat\subseteq \integers\subseteq \dyadic\subseteq
\rational\subseteq \algebraic\subseteq \ptcomplex\subseteq \complex$.

Let $\p$ ($\e$, resp.) be the class of all sets recognized by
certain polynomial-time (linear exponential-time, resp.) deterministic
TMs. Moreover, $\np$ is the class of all sets recognized by
polynomial-time nondeterministic TMs.  The class $\bpp$ ($\pp$, resp.)
denotes the class of all sets recognized by polynomial-time
probabilistic TMs with bounded-error (unbounded-error, resp.)
probability.

A function mapping from $\Sigma^*$ to $\Sigma^*$ is in $\fp$ if its
values are computed by a certain polynomial-time deterministic TM with
an output tape. A function $f$ from $\Sigma^*$ to $\nat$ is in
$\sharpp$ if there exists a polynomial-time nondeterministic TM $M$
such that $f(x)=\#M(x)$ for every $x\in\Sigma^*$ \cite{Val79}. By
expanding $\sharpp$ naturally, we define $\sharpe$, based on
$2^{O(n)}$-time nondeterministic TMs.  A function $f$ from $\Sigma^*$
to $\integers$ is in $\gapp$ if there exists a polynomial-time
nondeterministic TM $M$ such that $f(x)=\#M(x) - \#\overline{M}(x)$
for every $x\in\Sigma^*$ \cite{FFK94}. The class $\npsv$ is the
collection of all partial functions $f$ from $\Sigma^*$ to $\Sigma^*$
(called {\em single-valued NP-functions}) such that there exists a
polynomial-time nondeterministic TM $M$ with an output tape satisfying
the following: for every $x$, (i) if $x\in\dom(f)$ then $M$ on input
$x$ has at least one accepting computation path and all accepting
computation paths output precisely $f(x)$ and (ii) if
$x\not\in\dom(f)$ then all computation paths of $M$ on $x$ end with
rejecting configurations \cite{BLS84,SMB83}.

The class $\cequalp$ is the collection of all sets $A$ of the form
$A=\{x\in\Sigma^* \mid f(x)=0\}$ for certain $\gapp$-functions
$f$. The collections of all sets $A$ whose characteristic functions
$\chi_{A}$ belong to $\sharpp$ and $\gapp$ are respectively denoted
$\up$ and $\spp$.

An {\em oracle TM} induces relativization. 
In particular, an oracle TM is said
to make {\em nonadaptive queries} if, on every input $x$, $M$ makes a
list (called a {\em query list}) of strings that are all queried after
$M$ completes the list. The function class $\fp$ naturally induces the
{\em nonadaptive relativization} $\fp_{\|}^{A}$ (the {\em adaptive 
relativization} $\fp^{A}$, resp.) as the collection of
all functions computed by polynomial-time deterministic oracle TMs
that make nonadaptive (adaptive, resp.) 
queries to oracle $A$. Let $\CC$ be any
adaptively relativizable class of functions or sets.  A set $A$ is
called a {\em low set for $\CC$} if $\CC^A\subseteq \CC$, and the
notation $\low{\CC}$ denotes the class of all low sets for $\CC$. If
$\CC$ admits its nonadaptive relativization $\CC_{\|}^{(\cdot)}$, we
denote by $\low{\CC_{\|}}$ the class of all sets $A$ satisfying
$\CC_{\|}^A\subseteq \CC$.

A {\em pairing function} $\pair{\cdot,\cdot}$ is assumed to be
one-to-one on $\Sigma^*$ and polynomial-time computable with
polynomial-time computable inverses. For simplicity, we assume the
extra condition: $|\pair{x,y}|=r(1^{|x|+|y|})$ for all pairs $(x,y)$,
where $r$ is a certain fixed $\fp$-function.

For other standard notions and notation in classical complexity
theory, the reader should refer to recent textbooks, \eg
\cite{BDG88,DK00,HO02}.

\section{Quantum Turing Machines}\label{sec:QTM}

The notion of a {\em QTM} was originally introduced in {\cite{Deu85}}
and fully developed by a series of expositions
{\cite{BV97,NO02,ON99,Yam99}}. For convenience, we use in this paper a
general definition of QTMs\footnote{It is proven in
\cite{NO02,Yam99,Yao93} that our model is polynomially ``equivalent''
to the more restrictive model used in \cite{BV97}, which is sometimes
called {\em conservative} \cite{Yam99,YY99}.} given in {\cite{Yam99}},
where the QTM has $k$ two-way infinite tapes of cells indexed by
$\integers$ and its read/write heads move along the tapes either to
the left or to the right or the heads stay still. This model greatly
simplifies the programming of QTMs. This section gives basic notions
and notation associated with QTMs.

\subsection{Definition of Multiple Tape Quantum Turing Machines}
\label{sec:def-QTM}	

A {\em pure quantum state} is a unit-norm vector in a Hilbert space
(that is, a complex vector space with the standard inner product
$\measure{\cdot}{\cdot}$), where the {\em norm} $\|\qubit{\phi}\|$ of
a vector $\qubit{\phi}$ is defined as $\sqrt{\measure{\phi}{\phi}}$.
A {\em quantum bit} ({\em qubit}, for short) is a pure quantum state
in a $2$-dimensional Hilbert space. We often use the standard
computational basis $\{\qubit{0},\qubit{1}\}$ to represent a qubit.  A
{\em quantum string} ({\em qustring}, for short) {\em of size $n$} is
a pure quantum state in a Hilbert space of dimension
$2^n$. Thus, a qubit is a qustring of size $1$. The size of qustring
$\qubit{\phi}$ is denoted $\ell(\qubit{\phi})$. For each
$n\in\nat^{+}$, let $\Phi_{n}$ denote the collection of all qustrings
of size $n$ and set $\Phi_{\infty}=\bigcup_{n\in\nat^{+}}\Phi_{n}$.

There are four useful unitary transformations used in this paper. For
every angle $\theta\in[0,2\pi)$, the {\em phase shift} $P_{\theta}$
maps $\qubit{0}$ to $\qubit{0}$ and $\qubit{1}$ to
$e^{i\theta}\qubit{1}$.  The {\em Walsh-Hadamard transformation} $H$
changes $\qubit{0}$ into $\frac{1}{\sqrt{2}}(\qubit{0}+\qubit{1})$ and
$\qubit{1}$ into $\frac{1}{\sqrt{2}}(\qubit{0}-\qubit{1})$. The {\em
quantum Fourier transformation} $QFT_{n}$ maps $\qubit{m}$ to
$\frac{1}{\sqrt{2^n}}\sum_{\ell=0}^{2^n-1}e^{\frac{2\pi i
m\ell}{2^n}}\qubit{\ell}$, where we identify an integer between $0$
and $2^n-1$ with a binary string of length $n$ (in the lexicographic
order). The transformation $H_2$ acts on
$\{\qubit{\bar{0}},\qubit{\bar{1}},\qubit{\bar{2}},\qubit{\bar{3}}\}$
exactly as $H\otimes H$ acts on $\{\qubit{0},\qubit{1}\}^2$ by way of
identifying $\bar{0}=00$, $\bar{1}=01$, $\bar{2}=10$, and
$\bar{3}=11$.

Formally, a $k$-tape QTM $M$ is defined as a sextuple $(Q,q_0,Q_f,
\tilde{\Sigma}_k,\tilde{\Gamma}_k,\delta)$, where 
$\tilde{\Sigma}_{k}=
\Sigma_1\times\Sigma_2\times\cdots\times\Sigma_k$,
$\tilde{\Gamma}_{k}=
\Gamma_1\times\Gamma_2\times\cdots\times\Gamma_k$, each $\Sigma_i$ is
a finite (possibly empty) 
input/output alphabet for tape $i$, $\Gamma_i$ is a finite
tape alphabet for tape $i$ including $\Sigma_i$ as well as a
distinguished blank symbol $\#$, $Q$ is a finite set of (internal)
states including an initial state $q_0$, $Q_f$ is a nonempty set of
final states with $q_0\not\in Q_f\subseteq Q$, and $\delta$ is a total
multi-valued quantum transition function mapping from
$Q\times\tilde{\Gamma}_k$ to $\complex^{Q\times\tilde{\Gamma}_k
\times\{L,N,R\}^k}$. Note that each value $\delta(p,\vsigma)$ is 
described as a linear combination of the form $\sum
\alpha^{(p,\svsigma)}_{q,\svtau,\svd}
\qubit{q}\qubit{\vtau}\qubit{\vd}$, where the sum is taken over all
$q\in Q$, $\vd\in\{L,N,R\}^k$, and $\vsigma,\vtau\in\tilde{\Sigma}_k$,
and each complex number $\alpha^{(p,\svsigma)}_{q,\svtau,\svd}$ is
called an {\em amplitude} of $M$, which is also written as
$\delta(p,\vsigma,q,\vtau,\vd)$.  This $\delta$ induces the {\em
time-evolution operator} (or {\em matrix}), denoted $U_{M}$, which is
a unitary operator conducting a single application of $\delta$ to the
space spanned by all configurations of $M$ (called the {\em
configuration space} of $M$), where a {\em configuration} of $M$ is a
classical description of an internal state, all head positions, and
all tape contents.  In particular, the {\em initial configuration} of
$M$ on input $\vx\in\tilde{\Sigma}_k$ is a unique configuration in
which machine's state is $q_0$, every head rests on its start cell,
the input tapes contain $\vx$, and all other tapes are empty. A {\em
final configuration} of $M$ is a configuration of $M$ with a final
state. For language recognition, we define an {\em accepting
configuration} as a final configuration in which the output tape has
symbol ``1'' in its start cell. Any other final configurations are
simply called {\em rejecting configurations}. A {\em
computation path} of $M$ on input $x$ is a sequence of configurations
in which (i) the first configuration is the initial configuration of
$M$ on $x$ and (ii) any other configuration is obtained from its
predecessor by a single application of $M$'s transition function
$\delta$. A vector in the configuration space of $M$ is conventionally
called a {\em superposition} (of configurations) of $M$. In general, a
QTM can start with an arbitrary superposition, which is called an {\em
initial superposition}. We often restrict our interest on initial
superpositions that consist entirely of initial configurations with
string inputs of the same length so that we can identify such
superpositions with their inputs.

The {\em running time} of a QTM $M$ on input $\qubit{\phi}$ is defined
to be the minimal number $t$ (if any) such that all computation paths
of $M$ on $\qubit{\phi}$ simultaneously reach certain final
configurations at time $t$. We say that $M$ on input $\qubit{\phi}$
{\em halts at time $t$} ({\em within time $t$}, resp.) if its running
time is defined and is exactly $t$ (at most $t$, resp.).  We call $M$
a {\em polynomial-time QTM} if there exists a polynomial $p$ such
that, on every input $\qubit{\phi}\in\Phi_{\infty}$, $M$ halts exactly
at time $p(\ell(\qubit{\phi}))$.  This definition of polynomial-time
computation seems restrictive but it is easier to avoid the so-called
{\em timing problem}, which often arises when we modify QTMs (see, \eg
\cite{BV97,Oza02,Yam99} for detailed discussions). When $M$ halts on
input $\qubit{\phi}$, the superposition that is generated by $M$ on
$\qubit{\phi}$ is called the {\em final superposition of $M$ on}
$\qubit{\phi}$. For any superposition $\qubit{\phi}$ on which $M$
halts, the notation $M\qubit{\phi}$ denotes the final superposition of
$M$ that starts with $\qubit{\phi}$ as an initial superposition. By
linearity, $M\qubit{\phi} =\sum_{s\in CONF(M)}\alpha_s M\qubit{s}$ if
$\qubit{\phi}=\sum_{s\in CONF(M)}\alpha_s\qubit{s}$, where $CONF(M)$
is the set of all configurations of $M$ and each $\alpha_s$ is a
complex number.

The following terminology comes from \cite{BV97,Yam99}. For any
nonempty subset $K$ of $\complex$, we say that $M$ has {\em
$K$-amplitudes} if all amplitudes of $M$ are drawn from $K$. This $K$
is called an {\em amplitude set} of $M$. A QTM is {\em dynamic} if its
heads always move to the right or to the left (not staying still). A
dynamic QTM is {\em unidirectional} if, for any $p_1,p_2,q\in Q$,
$\vsigma_1,\vsigma_2\in\tilde{\Sigma}_k$, and
$\vd_1,\vd_2\in\{L,R\}^k$, $\delta(p_1,\vsigma_1,q,\vtau_1,\vd_1)\cdot
\delta(p_2,\vsigma_2,q,\vtau_2,\vd_2)\neq 0$ implies $\vd_1=\vd_2$. A
QTM $M$ is in {\em normal form} if, for every $q_f\in Q_f$, there
exists a vector $\vd\in\{L,N,R\}^k$ of directions such that
$\delta(q_f,\vsigma)=\qubit{q_0}\qubit{\vsigma}\qubit{\vd}$ for all
$\vsigma\in \tilde{\Gamma}_k$, and $M$ is {\em stationary} if, when it
halts, all heads halt in the start cells. Since a QTM $M$ may enter
final states several times before it halts, we need to call $M$ {\em
synchronous} if, for every qustring $\qubit{\phi}$, whenever any
computation path of $M$ on input $\qubit{\phi}$ enters a final state,
all computation paths of $M$ on $\qubit{\phi}$ enter (possibly
different) final states at the same time. A QTM $M$ is {\em
well-formed} if its time-evolution operator preserves the $L_2$-norm
(\ie $\|U_{M}\qubit{\phi}\|=\|\qubit{\phi}\|$ for all vectors
$\qubit{\phi}$ in the configuration space of $M$). We also use clean
QTMs, where a QTM $M$ is called {\em clean} if it is synchronous,
stationary, and in normal form and, when it halts, all tapes except
for the output tape become empty.

For any qustring $\qubit{\phi}$ and any string $y$, we say in general
that a well-formed QTM $M$ on input $\qubit{\phi}$ {\em outputs $y$
with probability $\alpha$} if $\alpha$ equals the sum of all squared
magnitudes of any configurations, in the final superposition of $M$ on
input $\qubit{\phi}$, in which the output tape consists only of
$\qubit{y}$ (where the leftmost symbol of $y$ is in the start
cell). Moreover, we say that $M$ {\em accepts} ({\em rejects}, resp.)
{\em $\qubit{\phi}$ with probability $\alpha$} if $\alpha$ equals the
sum of all squared magnitudes of any accepting (rejecting, resp.)
configurations in the final superposition of $M$ on input
$\qubit{\phi}$. The {\em acceptance probability} ({\em rejection
probability}, resp.) of $M$ on input $\qubit{\phi}$, denoted
$\rho_M(\qubit{\phi})$ ($\overline{\rho}_{M}(\qubit{\phi})$, resp.),
is the probability that $M$ accepts (rejects, resp.) input
$\qubit{\phi}$.  In particular, if $\qubit{\phi}$ is of the form
$\qubit{x}$ for classical string $x$, we omit the ket notation and
write, \eg $\rho_{M}(x)$ instead of $\rho_{M}(\qubit{x})$.

An {\em oracle QTM} is further equipped with a designated tape, called
a {\em query tape} and two distinguished states, a {\em pre-query
state} $q_p$ and a {\em post-query state} $q_a$. Let $A$ be any subset
of $\Sigma^*$ (called an {\em oracle}). The oracle QTM invokes an {\em
oracle query} (a {\em query}, for short) by entering state $q_p$ with
$\qubit{y}\qubit{b}$ written on the query tape, where $b\in\{0,1\}$
and $y\in\Sigma^*$. The leftmost symbol of $y$ is in the start
cell. In the special case where the query tape is empty, the machine
immediately enters $q_a$. In a single step, the tape content is
changed into $\qubit{y}\qubit{b\oplus \chi_{A}(y)}$ and the machine
enters state $q_a$ without moving any heads or altering 
any tape contents, where $\oplus$ denotes
the (bitwise) XOR.  The notation $M^{A}$ is used for oracle QTM $M$
with oracle $A$ and $\rho_{M}^A(x)$ denotes the acceptance probability
of $M^A$ on input $x$.

\paragraph{A Brief Discussion on QTMs.}
Firstly, the choice of amplitude set $K$ is crucial for most
applications of QTMs. Thus, we pay a special attention to the
amplitudes of a given QTM.  Throughout this paper, $K$ denotes an
arbitrary subset of $\complex$ that includes $\{0,\pm1\}$ for
convenience unless otherwise stated. All quantum function classes
discussed in this paper rely on the choice of amplitude set
$K$. Bernstein and Vazirani \cite{BV97} used $\ptcomplex$ as the basis
for their proof of the existence of a universal QTM. Although it is
debatable whether $\ptcomplex$ is the most natural choice of an
amplitude set for QTMs since $\algebraic$ is often used in many
quantum algorithms, we find it convenient in this paper to drop script
$K$ when $K=\ptcomplex$.

Secondly, we need to address the difference between a well-formed QTM
model and a model of a uniform family of quantum circuits. In many
proofs of this paper, we often give quantum-circuit descriptions, when
we define QTMs, instead of QTM descriptions. However, as was pointed
out in \cite{NO02}, these two models might not always define exactly
the same quantum complexity classes, particularly, $\eqp_{K}$ and
$\zqp_{K}$. Hence, whenever we give a quantum-circuit description, we
need to check if the actual implementation of a given quantum circuit
on a QTM is possible.

\subsection{Fundamental Lemmas}

For those who are not familiar with multi-tape QTMs with flexible head
moves, we first list six fundamental lemmas\footnote{Similar results
for the conservative QTMs are given in \cite{BBBV97,BV97}.}, given in
\cite{Yam99}, without their proofs. These lemmas will serve the later
sections.

The well-formedness of a QTM
$M=(Q,q_0,Q_f,\tilde{\Sigma}_k,\tilde{\Gamma}_k,\delta)$ is
characterized by the following three local conditions of its
transition function $\delta$.  Let $D=\{0,\pm1\}$,
$E=\{0,\pm1,\pm2\}$, and $H=\{0,\pm1,\natural\}$, where $\natural$ is
a distinguished symbol not in $\{0,\pm1\}$. For any
$\vepsilon=(\epsilon_i)_{1\leq i\leq k} \in E^k$, let $D_{\svepsilon}
=\{\vd\in D^k\mid \forall
i\in\{1,\ldots,k\}(|2d_i-\epsilon_i|\leq1)\}$ and, for any
$\vd=(d_i)_{1\leq i\leq k} \in D^k$, let $E_{\svd} =\{\vepsilon\in
E^k\mid \forall i\in\{1,\ldots,k\}(|2d_i-\epsilon_i|\leq1)\}$.  For
any $(p,\vsigma,\vtau)\in Q\times
\tilde{\Sigma}_k\times\tilde{\Sigma}_k$ and any $\vepsilon\in E^k$, define 
$\delta[p,\vsigma,\vtau|\vepsilon] = \sum_{q\in Q} \sum_{\svd\in
D_{\svepsilon}} \delta(p,\vsigma,q,\vtau,\vd)|E_{\svd}|^{-1/2}
\qubit{q} \qubit{h_{\svd,\svepsilon}}$, where 
$h_{\svd,\svepsilon} =(h_{d_i,\epsilon_i})_{1\leq i\leq k}\in H^k$ is
defined as $h_{d_i,\epsilon_i}=2d_i-\epsilon_i$ if $\epsilon_i\neq0$
and $h_{d_i,\epsilon_i}=\natural$ otherwise.

\begin{lemma}\label{lemma:well-formedness}
{\rm (Well-Formedness Lemma)}\hs{1} Let $k$ be any positive integer. A
$k$-tape QTM $(Q,q_0,Q_f,\tilde{\Sigma}_k,\tilde{\Gamma}_k,\delta)$ is
well-formed iff the following three conditions hold.
\vs{-2}
\begin{enumerate}
\item (unit length) $\|\delta(p,\vsigma)\|=1$ for all 
$(p,\vsigma)\in Q\times\tilde{\Gamma}_k$.
\vs{-2}
\item (orthogonality) $\delta(p_1,\vsigma_1) 
\cdot\delta(p_2,\vsigma_2)=0$ for any distinct pairs
$(p_1,\vsigma_1), (p_2,\vsigma_2)\in Q\times\tilde{\Gamma}_k$.
\vs{-2}
\item (separability) $\delta[p_1,\vsigma_1,\vtau_1|\vepsilon]\cdot
\delta[p_2,\vsigma_2,\vtau_2|\vepsilon']=0$ for any distinct pair
$\vepsilon,\vepsilon'\in E^k$ and for any pair 
\linebreak
$(p_1,\vsigma_1,\vtau_1), (p_2,\vsigma_2,\vtau_2)\in Q\times 
\tilde{\Gamma}_k \times\tilde{\Gamma}_k$.
\end{enumerate}
\end{lemma}

Note that, for any given well-formed QTM $M$, we can freely add extra
tapes and extra states to $M$ without changing the behavior of $M$ by
idling the heads on the extra tapes and instructing the extra states
to ``do nothing.'' A machine obtained in such a way is called a {\em
simple expansion} of $M$. By expanding two given well-formed QTMs, we
can always assume that they share the same configuration space.

Another important lemma known as the Completion Lemma states that any
partially-defined QTM can be expanded to a standard QTM. This allows
us to describe an evolution of the machine's superpositions only for
configurations of specific interest. We say that amplitude set $K$ is
{\em admissible} if it is closed under the following operations:
addition, subtraction, multiplication, division, complex conjugation,
and square root. For instance, $\algebraic$, $\ptcomplex$, and
$\complex$ are all admissible.

\begin{lemma}\label{lemma:completion}
{\rm (Completion Lemma)}\hs{1} Let $K$ be any admissible
set.\footnote{Certain non-admissible sets, such as
$\{0,\pm1,\pm\frac{3}{5},\pm\frac{4}{5}\}$, can satisfy the Completion
Lemma.} For any $k$-tape polynomial-time $K$-amplitude QTM with a
partially-defined quantum transition function $\delta$ that satisfies
the three conditions given in the Well-Formedness Lemma,
there exists a $k$-tape polynomial-time well-formed $K$-amplitude QTM
$M'$, with the same state set and alphabets, whose transition function
$\delta'$ agrees with $\delta$ whenever $\delta$ is defined.
\end{lemma}

The Reversal Lemma asserts the existence of a QTM that reverses a
given QTM. Let $M_1$ and $M_2$ be two well-formed QTMs with the same
tape alphabet. Assume that $M_1$ has a single final state. For any
input $x$ on which $M_1$ halts, let $c_x$ and $\qubit{\phi_x}$ be the
initial configuration and the final superposition of $M_1$ on $x$,
respectively. We say that $M_2$ {\em reverses the computation of}
$M_1$ if, for any input $x$ on which $M_1$ halts, $M_2$ starts with
$\qubit{\phi_x}$ as its initial superposition and halts in a final
superposition consisting entirely of configuration $c_x$ with 
amplitude $1$ \cite{BV97}. The machine $M_2$ is called a {\em
reversing machine} of $M_1$. The notation $K^{*}$ denotes the set of
all complex conjugates $\gamma^*$ for any numbers $\gamma\in K$.

\begin{lemma}\label{lemma:reverse}
{\rm (Reversal Lemma)}\hs{1} Assume that $K^*\subseteq K$. Let $M$ be
any polynomial-time synchronous dynamic normal-form unidirectional
well-formed $K$-amplitude QTM with a single final state. There exists
another synchronous dynamic normal-form unidirectional well-formed
$K$-amplitude QTM $M_R$ that reverses the computation of $\tilde{M}$
with extra constant steps.
\end{lemma}

The Reversal Lemma yields another useful lemma, called the Squaring
Lemma.

\begin{lemma}{\rm (Squaring Lemma)}\hs{1}
Assume that $K^*\subseteq K$.  Let $M$ be any polynomial-time
synchronous dynamic normal-form unidirectional well-formed
$K$-amplitude QTM, with a single final state, which outputs
$b(x)\in\{0,1\}$ on each input $x$ with probability
$\rho_{M}(x)$. There exists a synchronous dynamic normal-form
unidirectional well-formed QTM $N$ with $K$-amplitudes such that $N$
on input $x$ produces with linear slowdown the final superposition
containing the final configuration, with nonnegative real amplitude
$\rho_{M}(x)$ (and thus, probability exactly $\rho_{M}(x)^2$), in
which $N$ is in a final state with $x$ written on the input tape,
$b(x)$ on the output tape, and empty elsewhere.
\end{lemma}

The next lemma guarantees that any time-bounded well-formed QTM can be
converted into another well-formed QTM that is practically usable as a
subroutine of other QTMs. For the lemma, we need the following
notions. Let $\vk=(k_1,k_2,\ldots,k_m)$ with $1\leq
k_1<k_2<\cdots<k_m\leq k$ for $k\in\nat^{+}$. For any $k$-tape QTM
$M$, a function $f$ from $CONF(M)$ to $CONF(M)$ is said to {\em
preserve contents of tapes $\vk$} if, for every configuration $s\in
CONF(M)$, the contents of tapes $\vk$ in $s$ is identical to those
of tapes $\vk$ in $f(s)$. Let $M$ and $M'$ be any two
well-formed QTMs. We say that {\em $M'$ simulates $M$ on input
$\qubit{\phi}$ over tapes $\vk$} if there exists a simple expansion
$M_{exp}$ of $M$ such that (i) $M_{exp}$ and $M'$ share the same
configuration space (thus, $CONF(M_{exp})=CONF(M')$), (ii) there
exists a one-to-one function $f$ from $CONF(M_{exp})$ to $CONF(M')$
such that (i$'$) there is a certain polynomial-time deterministic TM
that, starting with each configuration $s$, halts in configuration
$f(s)$, (ii$'$) $f$ preserves contents of tapes $\vk$ of $M_{exp}$, and
(iii$'$) for any configuration $s$ in the final superposition of
$M_{exp}$ on input $\qubit{\phi}$, the amplitude of configuration $s$
in the final superposition of $M_{exp}$ on input $\qubit{\phi}$ equals
that of configuration $f(s)$ in the final superposition of $M'$ on
input $\qubit{\phi}$.

A QTM $M$ with tapes $\vk$ is said to be {\em quasi-stationary on
tapes} $\vk$ if, when it halts, the heads of tapes $\vk$ move back to
their start cells, and $M$ is in {\em quasi-normal form on tapes}
$\vk$ if, for every $q_f\in Q_f$, there exists a direction $\vd$ of
the heads of tapes $\vk$ such that, whenever $M$ is in state $q_f$, in
a single step (i) $M$ enters state $q_0$, (ii) the heads of tapes
$\vk$ move in direction $\vd$, and (iii) the contents of tapes $\vk$
are not altered. When the designated tapes $\vk$ are clear from the
context, $M$ is briefly called a quasi-stationary quasi-normal-form
QTM.

\begin{lemma}\label{lemma:conversion}
Any polynomial-time well-formed QTM $M$ with $K$-amplitudes can be
simulated over designated tapes by a certain polynomial-time
synchronous well-formed QTM $M'$, with a single final state, which is
also quasi-stationary and in quasi-normal form on these designated
tapes. If $K$ is admissible, then $M'$ can be a synchronous dynamic
stationary unidirectional well-formed $K$-amplitudes QTM in normal
form with a single final state.
\end{lemma}

The following lemma shows that any well-formed oracle QTM can be
modified to a certain canonical form.  For any subset $A$ of
$\Sigma^*$, let $A'=\{y01^{m-|y|-2}\mid m\geq|y|+2,y\in A\}$.

\begin{lemma}{\rm (Canonical Form Lemma)}\hs{1}
Let $M$ be any polynomial-time well-formed oracle 
QTM with $K$-amplitudes.
Let $A$ be any oracle. There exists a polynomial-time well-formed
oracle QTM $N$ with $K$-amplitudes such that, for every $x$, (i)
$N^{A'}$ simulates $M^{A}$ on input $x$ over $M$'s tapes, 
(ii) the length of any query word is
exactly the same on all computation paths of $N^{A'}$ on input $x$, and
(iii) $N^{A'}$ makes exactly the same number of queries along each
computation path on input $x$.
\end{lemma}

Next, we prove two lemmas, which are related to upper bounds of
acceptance probabilities of QTMs. The first lemma is folklore but we
include its proof for completeness.

\begin{lemma}\label{lemma:probability}
Let $\qubit{\phi},\qubit{\psi}\in\Phi_{\infty}$ and let $M$ and $N$ be
any two well-formed QTMs with the same configuration space. If $M$
halts on input $\qubit{\phi}$ and $N$ halts on input $\qubit{\psi}$,
then $|\rho_{M}(\qubit{\phi})-\rho_{N}(\qubit{\psi})|\leq
\|M\qubit{\phi}-N\qubit{\psi}\|$.
\end{lemma}

\begin{proof}
Let $\qubit{\phi_0}$ and $\qubit{\psi_0}$ denote respectively the
initial superpositions of $M$ on input $\qubit{\phi}$ and of $N$ on
input $\qubit{\psi}$. Let $A$ and $R$ be the sets of all accepting
configurations and rejecting configurations, respectively, of $M$ on
any string input of length at most
$\max\{\ell(\qubit{\phi}),\ell(\qubit{\psi})\}$.  For convenience, let
$E=A\cup R$.  Assume that $M\ket{\phi_0}=\sum_{i\in E}\alpha_i\ket{i}$
and $N\ket{\psi_0}=\sum_{i\in E}\beta_i\ket{i}$.  We want to evaluate
the value $2|\rho_{M}(\ket{\phi}) - \rho_{N}(\ket{\psi})|$.  This term
equals $|\rho_{M}(\ket{\phi}) - \rho_{N}(\ket{\psi})| +
|\overline{\rho}_{M}(\ket{\phi}) - \overline{\rho}_{N}(\ket{\psi})|$,
which is at most $\sum_{i\in A}\left||\alpha_i|^2 - |\beta_i|^2\right|
+ \sum_{i\in R}\left||\alpha_i|^2 - |\beta_i|^2\right|$. Obviously,
this equals $\sum_{i\in E}
\left|(|\alpha_i|-|\beta_i|)(|\alpha_i|+|\beta_i|)\right|$, which is
bounded above by $\sum_{i\in E}|\alpha_i-\beta_i||\alpha_i| +
\sum_{i\in E}|\alpha_i-\beta_i||\beta_i|$ since 
$||\alpha_i|-|\beta_i||\leq|\alpha_i-\beta_i|$. We obtain $\sum_{i\in
E}|\alpha_i-\beta_i||\alpha_i| \leq \left(\sum_{i\in E}|\alpha_i -
\beta_i|^2
\sum_{i\in E}|\alpha_i|^2\right)^{1/2} = 
\left(\sum_{i\in E}|\alpha_i - \beta_i|^2\right)^{1/2}$ by the 
Cauchy-Schwartz inequality. Similarly, $\sum_{i\in
E}|\alpha_i-\beta_i||\beta_i| \leq \left(\sum_{i\in E}|\alpha_i -
\beta_i|^2\right)^{1/2}$. Hence, $2|\rho_{M}(\ket{\phi}) -
\rho_{N}(\ket{\psi})|$ is bounded above by $2\left(\sum_{i\in
E}|\alpha_i - \beta_i|^2\right)^{1/2}$, which equals
$2\|M\qubit{\phi}-N\qubit{\psi}\|$.  
\end{proof}

Let $M$ be any well-formed oracle QTM, $A$ any oracle, and
$\qubit{\phi}$ any qustring. Let $q^t_y(M,A,\qubit{\phi})$ denote the
{\em query magnitude of string $y$ of $M^A$ on input $\qubit{\phi}$ at
time $t$}, which is defined as the sum of squared magnitudes in the
superposition of configurations $cf$ of $M^A$ on input $\qubit{\phi}$
at time $t$ such that $cf$ is in a pre-query state with query word $y$
\cite{BBBV97}. In particular, $q^0_y(M,A,\qubit{\phi})=0$. The
following is derived from a key lemma in \cite{BBBV97}.

\begin{lemma}\label{lemma:query-magnitude}
Let $M$ be a well-formed oracle QTM whose running time $t(n)$ does not
depend on the choice of oracles. For any two oracles $A$ and $B$ and
for any two qustrings $\qubit{\phi}$ and $\qubit{\psi}$ of size $n$,
\[
|\rho^A_M(\qubit{\phi})-\rho^B_M(\qubit{\psi})|\leq
\|\qubit{\phi}-\qubit{\psi}\| +
2\sqrt{t(n)}\left(\sum_{i=1}^{t(n)-1}\sum_{y\in A\triangle B}
q^i_y(M,A,\qubit{\phi})\right)^{1/2}.
\]
\end{lemma}

\begin{proof}
Let $\ket{\phi_0}$ and $\ket{\psi_0}$ be respectively the initial
superpositions of $M$ on input $\qubit{\phi}$ and on input
$\qubit{\psi}$. Note that
$\|\qubit{\phi_0}-\qubit{\psi_0}\|=\|\qubit{\phi}-\qubit{\psi}\|$. Let
$U_A$ and $U_B$ be the time-evolution operators of $M^A$ and $M^B$,
respectively. For each $i\in\{0,1,\ldots,t(n)\}$, let
$\ket{\phi_{i+1}}=U_{A}\ket{\phi_{i}}$,
$\ket{\psi_{i+1}}=U_{B}\ket{\psi_{i}}$, and
$\ket{E_i}=U_{A}\ket{\phi_i} - U_{B}\ket{\phi_i}$.  Note that
$\ket{\phi_{t(n)}} = U^{t(n)}_{B}\ket{\phi_0} +
\sum_{i=0}^{t(n)-1}U_{B}^{t(n)-i-1}\ket{E_i}$.  Thus,
$\|\ket{\phi_{t(n)}} - \ket{\psi_{t(n)}}\|$ equals
$\|U^{t(n)}_{B}(\ket{\phi_0} - \ket{\psi_0}) +
\sum_{i=0}^{t(n)-1}U_{B}^{t(n)-i-1}\ket{E_i}\|$, which is at most
$\|U^{t(n)}_{B}(\ket{\phi_0} - \ket{\psi_0})\| +
\sum_{i=0}^{t(n)-1}\|U_{B}^{t(n)-i-1}\ket{E_i}\|$. This term equals
$\|\ket{\phi_0} - \ket{\psi_0}\| + \sum_{i=0}^{t(n)-1}\|\ket{E_i}\|$
since $U_{B}$ is unitary.  The Cauchy-Schwartz inequality implies that
$\sum_{i=0}^{t(n)-1}\|\ket{E_i}\| \leq
\sqrt{t(n)}\left(\sum_{i=0}^{t(n)-1} \|\ket{E_i}\|^2\right)^{1/2}$.
Since $\qubit{E_i}$ depends only on the configurations, in
$\ket{\phi_i}$, in which $M$ is in a pre-query state with query words
in $A \triangle B$, we have $\|\ket{E_i}\|^2 \leq 4\sum_{y\in
A\triangle B}q^i_y(M,A,\ket{\phi})$, and thus
$\sum_{i=0}^{t(n)-1}\|\ket{E_i}\|^2 \leq 4
\sum_{i=1}^{t(n)-1}\sum_{y\in A\triangle B}q^i_y(M,A,\ket{\phi})$.
Since Lemma \ref{lemma:probability} relativizes,  we obtain
$|\rho^A_M(\ket{\phi}) -
\rho^B_M(\ket{\psi})| \leq \|\ket{\phi_{t(n)}} - \ket{\psi_{t(n)}}\|$.
The desired result therefore follows.
\end{proof}

\section{Quantum Functions with Classical Inputs}
\label{sec:quantum-function}

Over the past few decades, the function classes $\fp$, $\npsv$,
$\sharpp$, and $\gapp$ have played a major role in classical
complexity theory. Many old complexity classes have been redefined in
terms of these function classes. For example, any $\np$-set $S$ is
characterized simply by $S=\{x\mid f(x)>0\}$ for a certain
$\sharpp$-function $f$. Similarly, any $\pp$-set $S$ is written as
$S=\{x\mid g(x)>0\}$ for a certain $\gapp$-function $g$.  Fenner,
Fortnow, and Kurtz \cite{FFK94} further studied the extended notion of
gap-definable complexity classes. These function classes continue to
fascinate complexity theoreticians.

Quantum functions naturally expand the classical framework of
computation with the help of quantum interference and quantum
entanglement. We pay special interest to classifying these quantum
functions and clarifying their roles in quantum complexity theory.  In
particular, we focus on quantum functions whose inputs are classical
binary strings.

This paper recognizes two categories of quantum functions. The first
category includes polynomial-time exact quantum functions,
polynomial-time bounded-error quantum functions, and single-valued
$\qma$-functions. These three types are the functional generalizations
of the language classes $\eqp$ \cite{BV97}, $\bqp$ \cite{BV97}, and
$\qma$ \cite{Kit99,Kni96}. The second category of quantum functions
includes polynomial-time quantum probability functions and
polynomial-time quantum probability gap functions, which are quantum
analogues of $\sharpp$ and $\gapp$.

\subsection{Quantum Computable Functions}

This section defines three types of quantum functions, mapping from
$\Sigma^*$ to $\Sigma^*$, whose outcomes are computed by
polynomial-time well-formed QTMs with high probability. We first
recall the language classes $\eqp_{K}$, $\bqp_{K}$, and
$\qma_{K}$. Earlier, Bernstein and Vazirani \cite{BV97} introduced two
important complexity classes, $\eqp_{K}$ ({\em exact QP}) and
$\bqp_{K}$ ({\em bounded-error QP}), which are the collections of all
sets recognized by polynomial-time well-formed $K$-amplitude QTMs with
certainty and with bounded-error probability, respectively. Later,
Knill \cite{Kni96} and Kitaev \cite{Kit99} studied a quantum analogue
of $\np$, named $\qma_{K}$ ({\em quantum Merlin-Arthur}) in
\cite{Wat00} (also called $\mathrm{BQNP}_{K}$ in \cite{Kit99}), which
is the collection of all sets $A$ that are characterized by
polynomial-time $K$-amplitude well-formed QTMs $M$ and polynomials $p$
as follows: for every $x$, (i) if $x\in A$ then $M$ accepts input
$\qubit{x}\qubit{\phi}$ with probability at least $3/4$ for a certain
qustring $\qubit{\phi}\in\Phi_{p(|x|)}$ and (ii) if $x\not\in A$ then
$M$ accepts $\qubit{x}\qubit{\phi}$ with probability at most $1/4$ for
all qustrings $\qubit{\phi}\in\Phi_{p(|x|)}$, where $\qubit{x}$ is
given on the first input tape and $\qubit{\phi}$ is given on the
second input tape. These language classes can be naturally expanded
into function classes.

We begin with the functional version of $\eqp$---the class of all
quantum functions whose values are obtained with certainty by the
measurement of the output tapes of polynomial-time well-formed
QTMs. We call them {\em polynomial-time exact quantum computable} in a
fashion similar to polynomial-time computable functions.

\begin{definition}\label{def:exact}
Let $\feqp_K$ be the set of {\em polynomial-time exact quantum
computable functions with $K$-amplitudes}\footnote{The class $\feqp$
was independently introduced in \cite{BD99}.}; that is, there exists a
polynomial-time well-formed QTM with $K$-amplitudes such that, on
every input $x$, $M$ outputs $f(x)$ with probability $1$. In this
case, we say that $M$ {\em computes $f$ with certainty}.
\end{definition}

Next, we introduce another important function class $\fbqp_{K}$ that
is induced naturally from $\bqp_{K}$.

\begin{definition}\label{def:FBQP}
A function $f$ is {\em polynomial-time bounded-error quantum
computable with $K$-amplitudes} if there exists a polynomial-time
well-formed $K$-amplitude QTM $M$ that, on every input $x$, outputs
$f(x)$ with probability at least $3/4$. In this case, we say that $M$
{\em computes $f$ with bounded-error probability}. Let $\fbqp_{K}$
denote the set of all polynomial-time bounded-error quantum functions
with $K$-amplitudes.
\end{definition}

The success probability $3/4$ of $M$ in Definition \ref{def:FBQP} can
be amplified to $1-2^{-q(n)}$ for any fixed polynomial $q$. Lemma
\ref{lemma:conversion} implies that $M$ can be simulated over all
tapes of $M$ by a certain synchronous well-formed quasi-stationary
quasi-normal-form QTM $M'$ with a single final state. The
amplification is done by sequentially running $M'$ $6q(n)+1$ times in
a new blank area of the work tapes of $M'$ each time and then
outputting the majority values.
 
By Lemma \ref{lemma:reversible}, any deterministic computation can be
simulated by a certain reversible computation with polynomial
slowdown. Thus, we have $\fp\subseteq\feqp_K$ for any amplitude set
$K$ ($\supseteq\{0,\pm1\}$).

\begin{lemma}
$\fp\subseteq \feqp_{K} \subseteq \fbqp_{K}$.
\end{lemma}

The classes $\feqp_{K}$ and $\fbqp_{K}$ are respectively the
functional expansions of $\eqp_{K}$ and $\bqp_{K}$ in the following
sense: a set is in $\eqp_{K}$ ($\bqp_{K}$, resp.) iff its
characteristic function is in $\feqp_{K}$ ($\fbqp_{K}$, resp.). Thus,
well-known properties of $\eqp_{K}$ and $\bqp_{K}$ can be used to
derive the fundamental properties of $\feqp_{K}$ and $\fbqp_{K}$.

The class $\bqp_{K}$ is known to be robust with the choice of
amplitude set $K$. It is shown in \cite{ADH97,Kit97,Sho96} that
$\bqp_{\ptcomplex}=\bqp_{\rational} =
\bqp_{\{0,\pm1,\pm\frac{1}{\sqrt{2}}\}}$. This is easily translated
into function class: $\fbqp_{\ptcomplex} =\fbqp_{\rational} =
\fbqp_{\{0,\pm1,\pm\frac{1}{\sqrt{2}}\}}$.  Unlike $\bqp_{K}$,
$\eqp_{K}$ is sensitive to its underlying amplitude set $K$. For
instance, Adleman, DeMarrais, and Huang \cite{ADH97} showed that
$\eqp_{\complex}=\eqp_{\ptcomplex}= \eqp_{\algebraic\cap\real}$ while
Nishimura \cite{Nis02} proved that
$\eqp_{\{0,\pm1,\pm\frac{3}{5},\pm\frac{4}{5}\}}$ collapses to
$\p$. These results imply that $\feqp_{\complex}= \feqp_{\ptcomplex}
=\feqp_{\algebraic}$ and
$\feqp_{\{0,\pm1,\pm\frac{3}{5},\pm\frac{4}{5}\}}=\fp$.

As noted in \S\ref{sec:notions}, we drop subscript $K$ when
$K=\ptcomplex$ and write $\feqp$ for $\feqp_K$ and $\fbqp$ for
$\fbqp_{K}$. A simple example of an $\fbqp$-function is the Integer
Factorization Problem. Since Shor's quantum algorithm \cite{Sho97}
solves this problem in polynomial time with bounded-error probability,
it belongs to $\fbqp$. However, it is not yet known whether it falls
into $\feqp$.

Now, we consider the functional expansion of $\qma_{K}$. Similar to
$\npsv$, we define $\qmasv_{K}$ from $\qma_{K}$.

\begin{definition}\label{def:QMASV}
A partial function $f$ from $\Sigma^*$ to $\Sigma^*$ is called a {\em
single-valued QMA function with $K$-amplitudes} if there exist a
polynomial $p$ and a polynomial-time well-formed QTM $M$ with
$K$-amplitudes such that, for every string $x$, (i) if $x\in\dom(f)$
then $M$ on input $\qubit{x}\qubit{\phi}$ outputs
$\qubit{1}\qubit{f(x)}$ with probability at least $3/4$ for a certain
qustring $\qubit{\phi}$ of size $p(|x|)$ and, for every string
$y\in\Sigma^*-\{f(x)\}$ and every qustring $\qubit{\psi}$ of size
$p(|x|)$, $M$ on input $\qubit{x}\qubit{\psi}$ outputs
$\qubit{1}\qubit{y}$ with probability at most $1/4$ and (ii) if
$x\not\in\dom(f)$ then, for every string $y\in\Sigma^*$ and every
qustring $\qubit{\psi}$ of size $p(|x|)$, $M$ on input
$\qubit{x}\qubit{\psi}$ outputs $\qubit{1}\qubit{y}$ with probability
at most $1/4$. The first qubit $\qubit{1}$ in $\qubit{1}\qubit{y}$
indicates an accepting configuration. Let $\qmasv_{K}$ be the set of
all single-valued $\qma$-functions with $K$-amplitudes.
\end{definition}

Similar to $\fbqp$, we can amplify the success probability of the QTM
$M$ in Definition \ref{def:QMASV} from $3/4$ to $1-2^{-q(n)}$ in the
following fashion: by Lemma \ref{lemma:conversion}, $M$ is simulated
over its tapes by a certain synchronous well-formed
quasi-stationary quasi-normal-form QTM with a single final
state. Given input $x\in\Sigma^n$ together with qustring
$\qubit{\phi}\in\Phi_{mp(n)}$ that is sectioned into $m$ blocks of
equal $p(n)$ bits, where $m=6q(n)+1$, we run this new QTM $m$ times
sequentially using a new block each time and output the majority value
at the end. This procedure works because any entanglement of two or
more blocks does not increase the error probability of each run.
 
A relationship between $\qmasv_{K}$ and $\qma_{K}$ is described as
follows.

\begin{lemma}
$\mathrm{dom}(\qmasv_K)=\qma_{K}$.
\end{lemma}

\begin{proof}
For any set $A\in\qma_{K}$, let $f(x)=1$ if $x\in A$ and let $f(x)$ be
undefined otherwise. This $f$ satisfies that $A=\dom(f)$. It is easy
to show that $f$ is in $\qmasv_{K}$ by using the QTM that witnesses
$A$. Conversely, assume that $f\in\qmasv_{K}$ witnessed by a certain
polynomial-time well-formed QTM $M$. By our definition of accepting
and rejecting configurations, the same QTM $M$ witnesses
$\dom(f)$. Thus, $\dom(f)\in\qma_{K}$.  
\end{proof}

To treat partial functions in accordance with the previously-defined
total functions, we sometimes view the function class $\fbqp$ as a
class of partial single-valued functions: a partial function $f$ is in
$\fbqp_{K}$ if there exists a polynomial-time well-formed QTM $M$ such
that, for every string $x$, (i) if $x\in\dom(f)$ then $M$ on input $x$
outputs $\qubit{1}\qubit{f(x)}$ with probability at least $3/4$ and
(ii) if $x\not\in\dom(f)$ then, for every $y\in\Sigma^*$, $M$ on $x$
outputs $\qubit{1}\qubit{y}$ with probability at most $1/4$.

We obtain the following lemma.

\begin{lemma}
$\npsv\cup\fbqp_{K}\subseteq \qmasv_{K}$ if
$K\supseteq\{0,\pm1,\pm\frac{1}{\sqrt{2}}\}$.
\end{lemma}

\begin{proof}
Clearly, $\fbqp_{K}\subseteq\qmasv_{K}$ even though $\fbqp_{K}$ is
viewed as a class of partial functions. Let $f\in\npsv$. We can design
a polynomial-time deterministic TM $M$ with an appropriate polynomial
$p$ that satisfies the following conditions: for every string $x$, (i)
if $x\in\dom(f)$ then $M$ on input $\pair{x,y}$ outputs either $1f(x)$
or $0$ for all strings $y\in\Sigma^{p(|x|)}$ and there exists a string
$y_x\in\Sigma^{p(|x|)}$ such that $M$ on input $\pair{x,y_x}$ outputs
$1f(x)$ and (ii) if $x\not\in\dom(f)$ then $M$ on input $\pair{x,y}$
outputs $0$ for all strings $y\in\Sigma^{p(|x|)}$. Lemma
\ref{lemma:reversible} guarantees the existence of a reversible TM
$M'$ that simulates $M$. Note that the running time of $M'$ depends
only on the length of input.  Consider the QTM $N$ that carries out
the following algorithm.
\begin{quote}
{\sf On input $\qubit{x}\qubit{\phi}$, where $x\in\Sigma^n$ is given
on the first tape and $\qubit{\phi}\in\Phi_{p(n)}$ is on the second
input tape, Observe the second input tape. If $\qubit{y}$ is observed,
copy $y$ into a storage tape to avoid any future interference.
Simulate $M'$ on input $\pair{x,y}$. }
\end{quote}
Since any reversible TM can be simulated on a certain well-formed QTM,
$N$ is a polynomial-time well-formed QTM. Clearly, $N$ has
$K$-amplitudes. Therefore, $f$ belongs to $\qmasv_{K}$.  \end{proof}

A function class $\FF$ is said to be {\em closed under composition}
if, for every pair $f,g\in\FF$, $f\circ g$ is also in $\FF$, where
$f\circ g=\lambda x.f(g(x))$. We claim that $\feqp$, $\fbqp$, and
$\qmasv$ are all closed under composition. The proof is not difficult
and left to the avid reader.

\begin{lemma}
$\feqp$, $\fbqp$, and $\qmasv$ are all closed under composition.
\end{lemma}

\subsection{Quantum Probability Functions}

In classical complexity theory, the number of accepting computation
paths of a nondeterministic TM is a key to many complexity classes
known as {\em counting classes}, which include $\up$, $\np$,
$\cequalp$, $\spp$, and $\pp$. In the late 1970s, Valiant \cite{Val79}
introduced the class of functions that output such numbers, and coined
the name $\sharpp$ for this function class.  The class $\sharpp$ has
then become an important subject in connection to counting classes
(see, \eg a survey \cite{For97}).

The acceptance probability of a quantum computation plays a crucial
role similar to the number of accepting computation paths in a
classical computation. To study the behaviors of the acceptance
probabilities of a well-formed QTM, we need to consider quantum
functions that output such probabilities.  We briefly call these
functions {\em quantum probability functions}. Following Valiant's
notation $\sharpp$, we coin the new name $\sharpqp_K$ ({\em sharp QP})
for the class of all polynomial-time quantum probability
functions. This class greatly expands our scope of quantum functions.

Recall that $\rho_{M}(x)$ denotes the acceptance probability of
well-formed QTM $M$ on input $x$.

\begin{definition}\label{def:prob-function}
A function $f$ from $\Sigma^*$ to $[0,1]$ is called a {\em
polynomial-time quantum probability function with $K$-amplitudes} if
there exists a polynomial-time well-formed QTM $M$ with $K$-amplitudes
such that $f(x)= \rho_M(x)$ for all $x$.  In this case, we simply say
that $M$ {\em witnesses} $f$.  The notation $\sharpqp_{K}$ denotes the
set of all polynomial-time quantum probability functions with
$K$-amplitudes.
\end{definition}

Similar to the roles of $\sharpp$, $\sharpqp_{K}$ can characterize
many existing quantum complexity classes. For example, $\sharpqp_{K}$
can be used to define $\eqp_{K}$ and $\bqp_{K}$. Another important
example is the language class $\nqp_{K}$ ({\em nondeterministic QP})
introduced by Adleman, DeMarrais, and Huang \cite{ADH97} as a quantum
analogue of $\np$.  Recently, it has been proven that
$\nqp_{\{0,\pm1,\pm\frac{3}{5},\pm\frac{4}{5}\}}= \nqp_{\complex} =
\co\cequalp$ \cite{FGHP99,FR99,YY99}. 
Using $\sharpqp_{K}$, $\nqp_{K}$ is characterized simply as the
collection of all sets of the form $\{x\mid f(x)>0\}$ for certain
$\sharpqp_{K}$-functions $f$.

If restricted to $\{0,1\}$-valued functions, quantum probability
functions coincide with exact computable functions. Recall that
$\{0,1\}^{\Sigma^*}$ denotes the set of all functions from $\Sigma^*$
to $\{0,1\}$.  Thus, $\feqp_{K}\cap \{0,1\}^{\Sigma^*} =
\sharpqp_{K}\cap
\{0,1\}^{\Sigma^*}$ for any amplitude set $K$.

The lemma below shows that $\sharpqp_{K}$ naturally expands $\sharpp$
if $K\supseteq \{0,\pm1,\pm\frac{1}{2}\}$.

\begin{lemma}\label{lemma:inclusion}
Let $K=\{0,\pm1,\pm\frac{1}{2}\}$. For every $\sharpp$-function $f$,
there exist two functions $\ell\in\fp\cap\nat^{\Sigma^*}$ and $g\in
\sharpqp_{K}$ such that $f(x)=\ell(1^{|x|})g(x)$ for every $x$.
\end{lemma}

\begin{proof}
In this proof, we use the tape alphabet
$\Gamma_{4}=\{\bar{0},\bar{1},\bar{2},\bar{3}\}$. Let $f$ be any
function in $\sharpp$. Take a polynomial $p$ and a polynomial-time
deterministic TM $M$ such that $f(x)=|\{y\in\Gamma_{4}^{p(|x|)}\mid
M(\pair{x,y})=1\}|$ for all $x$. {}From Lemma \ref{lemma:reversible},
we can assume that $M$ is reversible and its running time depends only
on the length of input.  Define the new QTM $N$ as follows.
\begin{quote}
{\sf On input $x$, write $\qubit{\bar{0}^{p(n)}}$ on a separate blank
tape and apply $H_2^{\otimes p(n)}$. Observe $\qubit{y}$ on this tape
and copy it into a storage tape to avoid any future
interference. Simulate $M$ on input $\pair{x,y}$.}
\end{quote}
Note that $N$ has $K$-amplitudes since the above procedure can be
conducted by a series of unitary operators with
$K$-amplitudes. Clearly, $\rho_{M}(x)$ equals $f(x)/4^{p(n)}$. It
suffices to set $\ell(x)=4^{p(|x|)}$ and $g(x)=\rho_{M}(x)$.  
\end{proof}

A similar argument of the above proof works for many other amplitude
sets $K$, including $\{0,\pm1,\pm\frac{3}{5},\pm\frac{4}{5}\}$.

The class $\sharpqp_{K}$ enjoys numerous closure properties.  To
describe these properties, we introduce the notion of {\em qubit
sources}. For any fixed function $\ell\in\nat^{\nat}$ and any index
set $I$, an ensemble $\{\qubit{\phi_x}\}_{x\in I}$ of qustrings is
called an {\em $\ell$-qubit source with $K$-amplitudes} if there
exists a polynomial-time well-formed clean $K$-amplitude QTM that, on
every input $x$, produces $\qubit{\phi_x}$ of size $\ell(|x|)$ on its
output tape.

\begin{lemma}\label{lemma:operators}
Let $f,g\in \sharpqp_K$, $p,h\in \feqp_K$ with $|p(1^n)|\in O(\log
n)$, and let $\ell$ be any polynomial.
\vs{-2}
\begin{enumerate}
\item $f\circ h\in\sharpqp_K$, where $f\circ h$ denotes the 
composition $\lambda x.f(h(x))$. 
\vs{-2}
\item If ensemble $\Phi=\{\qubit{\phi_x}\}_{x\in\{0,1\}^*}$ is an 
$\ell$-qubit source with $K$-amplitudes, then $\lambda
x.(\sum_{s:|s|=\ell(|x|)}|\measure{s}{\phi_{x}}|^2 f(\pair{x,s}))$ is
in $\sharpqp_K$.  In particular, if $K$ includes
$\{0,\pm1,\pm\frac{1}{2}\}$ then $\lambda x.\frac{1}{2}(f(x)+g(x))$ is
in $\sharpqp_{K}$.
\vs{-2}
\item $\lambda x.(\prod_{s:|s|=|p(1^{|x|})|} f(\pair{x,s}))$ is in 
$\sharpqp_K$. 
\vs{-2}
\item $\lambda x.f(x)^{|h(x)|}$ is in $\sharpqp_K$.
\end{enumerate}
\end{lemma}

\begin{proof}
1) Let $M_h$ be any polynomial-time well-formed $K$-amplitude QTM that
computes $h$ with certainty and let $M$ be any polynomial-time
well-formed $K$-amplitude QTM whose acceptance probability $\rho_{M}$
equals $f$. {}From Lemma \ref{lemma:conversion}, $M_h$ can be 
synchronous with a single final state as well as
quasi-stationary and in quasi-normal form on the output tape. Define
the new QTM $N$ as follows.
\begin{quote}
{\sf On input $x$, simulate $M_h$. Note that the head of the output
tape returns to the start cell. Observe the output tape after $M_h$
enters a unique final state. When $\qubit{y}$ is observed, simulate
$M$ on input $y$ using a new set of blank tapes. }
\end{quote}
Notice that the final superposition of $M_h$ must have the form
$\qubit{\psi}\qubit{h(x)}$, where $\qubit{h(x)}$ is the content of the
output tape. Since $\qubit{\psi}$ does not affect $M$'s move, the
acceptance probability $\rho_{N}(x)$ of $N$ is exactly
$\rho_{M}(h(x))$. Thus, we obtain $\rho_{N}(x)=f\circ h(x)$.

2) Let
$g(x)=\sum_{s:|s|=\ell(|x|)}|\measure{s}{\phi_x}|^2f(\pair{x,s})$ for
all $x$. Since $\{\qubit{\phi_x}\}_{x\in\Sigma^*}$ is an $\ell$-qubit
source with $K$-amplitudes, let $M_0$ be any polynomial-time
well-formed clean $K$-amplitude QTM that produces qustring
$\qubit{\phi_x}$ on input $x$. Let $M$ be another polynomial-time
well-formed $K$-amplitude QTM witnessing $f$. Consider the QTM $N$
that executes the following algorithm.
\begin{quote}
{\sf On input $x$, copy $x$ into a storage tape and then simulate
$M_0$ to produce $\qubit{\phi_x}$ on a new blank tape. Observe string
$s$ on this tape. Copy $s$ into a storage tape and then simulate $M$
on input $\pair{x,s}$.  }
\end{quote}
Obviously, $N$ has $K$-amplitudes. Note that the probability of
observing $s$ is exactly $|\measure{s}{\phi_x}|^2$. Note that copying
$s$ prevents any further interference between two computations of $M$
on input $\pair{x,s}$ and on different input $\pair{x,s'}$. Thus,
$\rho_{N}(x)= \sum_{s:|s|=\ell(|x|)}|\measure{s}{\phi_x}|^2
\rho_{M}(\pair{x,s})$, which implies $g(x)=\rho_{N}(x)$.

The second part follows from the fact that
$\{\frac{1}{2}\sum_{s:|s|=2}\qubit{s}\}_{x\in\Sigma^*}$ is $2$-qubit
source with $\{0,\pm1,\pm\frac{1}{2}\}$-amplitudes. In this case, we
define $f'$ as $f'(\pair{x,0b})= f(x)$ and $f'(\pair{x,1b}) = g(x)$
for each $b\in\{0,1\}$ and apply the first part.

3) Since $|p(1^n)|\in O(\log{n})$, there exists a constant $c\geq0$
such that $|p(1^n)|\leq c\log{n}+c$ for all $n\in\nat$. For a given
$f$, let $M$ be any polynomial-time well-formed $K$-amplitude QTM that
witnesses $f$. By Lemma \ref{lemma:conversion}, $M$ can be simulated
over its tapes by a certain polynomial-time synchronous well-formed
quasi-stationary quasi-normal-form $K$-amplitude QTM $M'$ with a
single final state. Consider the following QTM $N$.
\begin{quote}
{\sf On input $x$, compute $m=|p(1^{|x|})|$ deterministically. Write
$s:=0^{m}$ on a counter tape. Repeat the following by incrementing
lexicographically string $s$ written on the counter tape. Copy $x$ and
$s$ into a new blank area of a work tape and then simulate $M'$ on
input $\pair{x,s}$. If all runs of $M'$ end with accepting
configurations, then accept; otherwise, reject. }
\end{quote}
Obviously, each run of $M$ is independent because of the use of a new
blank area each time. Thus, the acceptance probability of $N$ equals
$\prod_{s:|s|=|p(1^{|x|})|}\rho_{M}(\pair{x,s})$.  Moreover, the
number of runs of $M$ on $x$ is exactly $2^{|p(1^{|x|})|}\leq 2^cn^c$,
which is polynomially bounded. Hence, $N$ runs in polynomial time.

4) Let $M_f$ be any well-formed $K$-amplitude QTM that witnesses $f$
in time polynomial $q$. Lemma \ref{lemma:conversion} yields the
existence of a polynomial-time synchronous well-formed
quasi-stationary quasi-normal-form $K$-amplitude QTM $M_f'$, with a
single final state, that simulates $M_f$ over all the tapes of
$M_f$. Since $h\in\feqp_{K}$, choose a polynomial $p$ that satisfies
$|h(x)|\leq p(|x|)$ for all $x$. Consider the following algorithm.
\begin{quote}
{\sf On input $x$, run $M_f'$ $|h(x)|$ times and idle $q(|x|)$ steps
$p(|x|)-|h(x)|$ times to avoid the timing problem.  Accept $x$ if all
the first $|h(x)|$ runs of $M_f'$ reach accepting configurations;
reject $x$ otherwise.  }
\end{quote}
Similar to 3), the above algorithm accepts $x$ with probability
exactly $\rho_{M_f}(x)^{|h(x)|}$ since $\rho_{M_f}(x)=\rho_{M_f'}(x)$.
\end{proof}

Although $\sharpqp$ enjoys useful closure properties as shown in Lemma
\ref{lemma:operators}, it is obviously not closed under subtraction. 
In classical context, Fenner, Fortnow, and Kurtz \cite{FFK94} studied
the subtraction closure of $\sharpp$, named $\gapp$. Similarly, the
subtraction closure of $\sharpqp$ can be naturally introduced.  We
call such functions {\em quantum probability gap functions}.  A
quantum probability gap function is formally defined to output the
difference between the acceptance and rejection probabilities of a
certain well-formed QTM.

\begin{definition}
A function $f$ from $\Sigma^*$ to $[-1,1]$ is called a {\em
polynomial-time quantum probability gap function with $K$-amplitudes}
if there exists a polynomial-time well-formed QTM $M$ with
$K$-amplitudes such that, for every $x$, $f(x)$ equals
$\rho_{M}(x)-\overline{\rho}_{M}(x)$. In other words,
$f(x)=2\rho_M(x)-1$ since $\rho_{M}(x)+\overline{\rho}_{M}(x)=1$.  Let
$\gapqp_{K}$ denote the set of all polynomial-time quantum probability
gap functions with $K$-amplitudes.
\end{definition}

Many closure properties of $\gapqp_{K}$ directly follow from those of
$\sharpqp_{K}$ using the closely-knitted relationships between
$\sharpqp_{K}$ and $\gapqp_{K}$ shown in \S\ref{sec:sharpqp-gapqp}.

\section{Relationships among Quantum Functions}

Empowered by quantum mechanism, quantum computation can draw close
quantum functions of different nature. The relationships among these
quantum functions are of special interest because they partly reveal
an essence of quantum computations. In the 1990s, several useful
techniques have been developed to analyze the behaviors of quantum
computations. These techniques are extensively used in this section to
show close connections among the quantum functions introduced in
\S\ref{sec:quantum-function}.

\subsection{A Relationship between \#QP and FBQP}

Stockmeyer \cite{Sto85} showed that every $\sharpp$-function can be
approximated deterministically with help of oracles chosen from
$\np^{\np}$. Later, Jerrum, Valiant, and Vazirani \cite{JVV86}
presented a randomized approximation scheme for $\sharpp$-functions
with an access to $\np$-oracles.  The approximation of
$\sharpqp$-functions is quite different. Obviously, if the range of a
$\sharpqp$-function $f$ is restricted to the set $\{0,1\}$, then $f$
falls into $\feqp$. For a general range, every $\sharpqp$-function can
be approximated quantumly without any help of oracles. We use an
amplitude amplification technique of Brassard, H{\o}yer, and Tapp
\cite{BHT98} to prove this claim.

In the mid 1990s, Grover \cite{Gro96} discovered a fast quantum
algorithm for a database search problem.  His algorithm is designed to
find the location of a target key word in a large database provided
that there is a unique location for the key word.  Brassard
\etalc~\cite{BHT98} elaborated Grover's database search algorithm and
showed how to compute with $\epsilon$-accuracy the norm of a given
superposition with high probability.

In what follows, we show that every $\sharpqp$-function can be
approximated by a certain $\fbqp$-function, where we view
$\fbqp$-functions as functions mapping from $\Sigma^*$ to $\dyadic$.

\begin{theorem}\label{theorem:sharpqp-fbqp}
$\sharpqp\appsubseteq\fbqp\cap\dyadic^{\Sigma^*}$.
\end{theorem}

\begin{proof}
Let $f$ be any $\sharpqp$-function, which is witnessed due to
Lemma \ref{lemma:conversion} by a certain polynomial-time synchronous
dynamic stationary unidirectional well-formed $\ptcomplex$-amplitude
QTM $M$ in normal form with a single final state. We also assume
without loss of generality that $M$ always outputs either $0$ or $1$
in the start cell.  By the Reversal Lemma, there exists the reversing
QTM $M_{R}$ of $M$. Let $q$ be any polynomial that bounds the running
times of both $M$ and $M_{R}$. For simplicity, assume that $M$ and
$M_R$ share the same configuration space.  Let $p$ be any positive
polynomial. Define $k(n)=\ceilings{\log p(n)}+3$ for all $n\in\nat$.
For convenience, we use integers between $0$ and $2^{k(n)}-1$ instead
of string of length $k(n)$.  By attaching a new blank storage tape to
$M$, we obtain the simple expansion of $M$, say $M'$. Note that $M'$
does not alter the content of this storage tape.

We define the quantum algorithm $\QQ_n$ that starts with any
superposition of $M'$ on input of length $n$. In this algorithm, we
check only the cells indexed between $-q(n)$ and $q(n)$. Let $I$
denote the identity operator.
\begin{quote}
{\sf Apply $-P_{\pi}$ to the bit written in the start cell of the
output tape of $M'$. Simulate $M_R$ on all the tapes except for the
storage tape. Check if (i) all the tapes except for the input and
storage tapes are blank and (ii) the contents of the input tape and
the storage tape agree. If so, apply $-I$; otherwise, apply $I$.
Finally, simulate $M$. }
\end{quote}
Let $x$ be any string of length $n$. For readability, we write $k$ for
$k(n)$. Consider the following two superpositions.  Let
$\qubit{\Phi^{(0)}}$ and $\qubit{\Phi^{(1)}}$ be respectively the
superpositions of
all final configurations of $M'$ in which $M'$ starts with input $x$
given to both the input and storage tapes and halts with
bit $0$ and with bit $1$ written on 
the output tape. Let $\theta$ be the
real number in $[0,\frac{\pi}{2}]$ satisfying that $\sin\theta =
\|\qubit{\Phi^{(1)}}\|$. Clearly, 
$f(x)=\sin^2\theta$. The algorithm $\QQ_n$ has two eigenvalues
$e^{2i\theta}$ and $e^{-2i\theta}$ with their corresponding
eigenvectors $\qubit{\Psi_0} =
\frac{1}{\sqrt{2}} 
\left(\frac{e^{i\theta}}{\cos\theta}\qubit{\Phi^{(0)}}
-\frac{ie^{i\theta}}{\sin\theta}\qubit{\Phi^{(1)}}\right)$ and
$\qubit{\Psi_1} =
\frac{1}{\sqrt{2}} 
\left(\frac{e^{-i\theta}}{\cos\theta}\qubit{\Phi^{(0)}}
+\frac{ie^{-i\theta}}{\sin\theta}\qubit{\Phi^{(1)}}\right)$; namely,
$\QQ\qubit{\Psi_0} = e^{2i\theta}\qubit{\Psi_0}$ and
$\QQ\qubit{\Psi_1} = e^{-2i\theta}\qubit{\Psi_1}$). 

To approximate $f(x)$, we need to estimate $\theta$. This is done by
the following phase estimation algorithm.  On input $x$ of length $n$,
copy $x$ into the storage tape to remember $x$ and then simulate $M$
on input $\qubit{x}$.  When $M$ halts, produce $\qubit{0^k}$ on a new
blank memory tape and apply $H^{\otimes k}$ to generate
$\frac{1}{\sqrt{2^k}}\sum_{m=0}^{2^k-1}\qubit{m}$.  Observe
$\qubit{m}$ and apply $\QQ_n$ $m$ times to all the tapes except for
this memory tape. Since $\qubit{\Phi^{(0)}}+\qubit{\Phi^{(1)}} =
\frac{1}{\sqrt{2}}(\qubit{\Psi_0}+\qubit{\Psi_1})$, we then obtain
qustring
$\frac{1}{\sqrt{2^k}}\sum_{m=0}^{2^k-1} 
\qubit{m}(e^{-2mi\theta}\qubit{\Psi_0}+e^{2mi\theta}\qubit{\Psi_1})$.
Next, we apply QFT$_{k}$ to the memory tape and observe this tape.
After QFT$_{k}$, the sum of the squared norms of both
$\qubit{\floors{\frac{2^k\theta}{\pi}}}\qubit{\Psi_0}$ and
$\qubit{\ceilings{\frac{2^k\theta}{\pi}}}\qubit{\Psi_0}$ becomes at
least $\frac{4}{\pi^2}$. Similarly, the squared norms of both
$\qubit{2^k-\floors{\frac{2^k\theta}{\pi}}}\qubit{\Psi_1}$ and
$\qubit{2^k-\ceilings{\frac{2^k\theta}{\pi}}}\qubit{\Psi_1}$ sum up to
at least $\frac{4}{\pi^2}$.  After observing $\qubit{\ell}$ on the
memory tape, we define $\ell'$ as $\ell'=\ell$ if
$\ell\leq\frac{2^k}{2}$ and $\ell'=2^k-\ell$ otherwise. The
probability that either $\ell'=\floors{\frac{2^k\theta}{\pi}}$ or
$\ell'=\ceilings{\frac{2^k\theta}{\pi}}$ is at least
$\frac{8}{\pi^2}$, which is larger than $\frac{4}{5}$. Moreover,
$|\theta-\frac{\pi\ell'}{2^k}|<\frac{\pi}{2^k}$ and thus,
$|\sin^2\theta -\sin^2\frac{\pi\ell'}{2^k}|\leq
\frac{\pi^2}{2^{2k}}<\frac{1}{4p(n)}$.  It follows from $f(x)
=\sin^2\theta$ that $|f(x) - \sin^2\frac{\pi
\ell'}{2^k}|<\frac{1}{4p(n)}$. At the end, we output a 
$\frac{1}{4p(n)}$-approximation of the value $\sin^2\frac{\pi
\ell'}{2^k}$.

Unfortunately, our algorithm has a minor problem due to the fact that
no QTM can carry out QFT$_{k}$ exactly \cite{NO02}. However, it is
possible to replace QFT$_{k}$ by a certain polynomial-time well-formed
QTM that $\frac{1}{4p(n)}$-approximates QFT$_{k}$. Therefore, $f$ is
in $\fbqp$.  
\end{proof}

Theorem \ref{theorem:sharpqp-fbqp} cannot be improved to
$\sharpqp\stappsubseteq\fbqp\cap\dyadic^{\Sigma^*}$ unless $\bqp=\pp$.

\begin{proposition}
If $\bqp\neq\pp$ then
$\sharpqp\not\stappsubseteq\fbqp\cap\dyadic^{\Sigma^*}$.
\end{proposition}

\begin{proof}
We show the contrapositive. Assume that
$\sharpqp\stappsubseteq\fbqp\cap\dyadic^{\Sigma^*}$. Let $A$ be any
set in $\pp$. There exists a $\sharpp$-function $f$ and a polynomial
$p$ such that, for every $x$, $x\in A$ implies
$f(x)>\frac{1}{2}+2^{-p(|x|)}$ and $x\not\in A$ implies $f(x)\leq
\frac{1}{2}$. By Lemma
\ref{lemma:inclusion}, there exist two functions $g\in\sharpqp$ and
$\ell\in\fp\cap\nat^{\Sigma^*}$ such that $f(x)=g(x)\ell(x)$ for all
$x$. Let $q$ be any polynomial satisfying $\ell(x)\leq 2^{q(|x|)}$ for
all $x$. Since $\sharpqp\stappsubseteq\fbqp\cap\dyadic^{\Sigma^*}$, we
can choose a function $h\in\fbqp\cap\dyadic^{\Sigma^*}$ such that
$|g(x)-h(x)|\leq 2^{-p(|x|)-q(|x|)-2}$ for all $x$. Let
$x\in\Sigma^n$. On one hand, if $x\in A$ then
$g(x)>\frac{1}{\ell(x)}(\frac{1}{2}+2^{-p(|x|)})\geq
\frac{1}{2\ell(x)}+2^{-p(n)-q(n)}$, which implies
$h(x)>\frac{1}{2\ell(x)}+2^{-p(n)-q(n)-1}$. On the other hand, if
$x\not\in A$ then $g(x)\leq \frac{1}{2\ell(x)}$. Thus,
$h(x)<\frac{1}{2\ell(x)}+2^{-p(n)-q(n)-1}$. Clearly, $h$ and $\ell$
can determine the membership of $A$. Since they are both in $\fbqp$,
$A$ belongs to $\bqp$. Therefore, $\pp\subseteq\bqp$. Since
$\bqp\subseteq\pp$, we obtain $\bqp=\pp$.  
\end{proof}

\subsection{Relationships between \#QP and GapQP}
\label{sec:sharpqp-gapqp}

We consider the relationship between $\gapqp$ and $\sharpqp$. Quantum
nature brings them closer than classical counterparts, $\gapp$ and
$\sharpp$. We first show that $\gapqp$ is the subtraction closure of
$\sharpqp$. More precisely, for any two sets $\FF$ and $\GG$ of
functions, let $\FF-\GG$ denote the set of all functions of the form
$f-g$, where $f\in\FF$ and $g\in\GG$. In Proposition
\ref{prop:gap-character}, we prove that $\gapqp=\sharpqp -
\sharpqp$. This shows another characterization of $\gapqp$.  As an
immediate consequence, we obtain $\sharpqp\subseteq\gapqp$.

Let $M$ be any synchronous QTM with $K$-amplitudes. If $M$ is
quasi-stationary on its unique output tape, let $\overline{M}$ denote
the QTM that simulates $M$ on all the tapes and, when $M$ halts, flips
$M$'s output bit. Note that if $M$ has $K$-amplitudes then
$\overline{M}$ also has $K$-amplitudes.

\begin{proposition}\label{prop:gap-character}
If $K\supseteq \{0,\pm1,\pm\frac{1}{2}\}$ then
$\gapqp_K=\sharpqp_K-\sharpqp_K$.
\end{proposition}

\begin{proof}
Let $f$ be any function in $\gapqp_{K}$. There exists a
polynomial-time well-formed QTM $M$ with $K$-amplitudes such that $f
=\rho_{M} - \overline{\rho}_{M}$. By Lemma \ref{lemma:conversion}, 
we assume that $M$
is synchronous and quasi-stationary on its output
tape. Note that $\overline{\rho}_{M}(x)$ coincides with
$\rho_{\overline{M}}(x)$ for every $x$. Thus, $f= \rho_{M} -
\rho_{\overline{M}}$. Since $\rho_{M}$ and $\rho_{\overline{M}}$ are
both $\sharpqp_{K}$-functions, $f$ belongs to
$\sharpqp_K-\sharpqp_K$. Thus, $\gapqp_{K}\subseteq \sharpqp_{K} -
\sharpqp_{K}$.

Conversely, assume that $f\in\sharpqp_{K} - \sharpqp_{K}$. It follows
from Lemma \ref{lemma:conversion} that there exist two polynomial-time
synchronous well-formed $K$-amplitude QTMs $M_g$ and $M_h$, which are
both quasi-stationary on their output tape, such that $f= \rho_{M_g} -
\rho_{M_h}$. Now, consider $\overline{M}_h$. Generally speaking, the
halting timing of $M_g$ may differ from that of $\overline{M}_h$. Let
$p_g$ and $p_h$ be polynomials that measure the running times of these
machines $M_g$ and $M_h$, respectively. Without loss of generality, we
can assume that $M_g$ and $M_h$ have the same alphabet, states, and
tapes.  To synchronize the halting timing of these machines, we attach
a {\em counter tape} that behaves like a clock (counting the number of
steps). When the machines halt, we force them to idle until the
counter hits $p_g(|x|)+p_h(|x|)$. Thus, we can assume that $M_g$ and
$\overline{M}_h$ halt at the same time. Now, consider the following
QTM $N$ whose tape alphabet includes
$\Gamma_4=\{\bar{0},\bar{1},\bar{2},\bar{3}\}$.
\begin{quote}
{\sf On input $x$, first write $\qubit{\bar{0}}$ on a separate blank
tape and apply $H_2$. Observe this tape. When either $\qubit{\bar{0}}$
or $\qubit{\bar{1}}$ is observed, simulate $M_g$ on input
$x$. Otherwise, simulate $\overline{M}_h$ on input $x$. }
\end{quote}
Note that $N$ has $K$-amplitudes since
$K\supseteq\{0,\pm1,\pm\frac{1}{2}\}$.  The acceptance probability
$\rho_{N}(x)$ is exactly $\frac{1}{2}\rho_{M_g}(x)+
\frac{1}{2}(1-\rho_{M_h}(x))$. Thus, the gap $2\rho_N(x)-1$ is exactly
$\rho_{M_g}(x)-\rho_{M_h}(x)$, which equals $f(x)$. Therefore,
$f\in\gapqp_{K}$. This concludes that
$\sharpqp_{K}-\sharpqp_{K}\subseteq \gapqp_{K}$.
\end{proof}

Since $\sharpqp_{K}\subseteq \sharpqp_{K} - \sharpqp_{K}$ for any $K$,
we obtain the following corollary.

\begin{corollary}
$\sharpqp_{K}\subseteq\gapqp_{K}$ if $K\supseteq
\{0,\pm1,\pm\frac{1}{2}\}$.
\end{corollary}

Unfortunately, it is unknown whether nonnegative $\gapqp$-functions
are all in $\sharpqp$. Here, we present only a partial solution to
this question. In the late 1990s, Fenner, Green, Homer, and Pruim
\cite{FGHP99} proved that, for every $\gapp$-function $f$, there
exists a polynomial-time well-formed QTM that accepts input $x$ with
probability $2^{-p(|x|)}f^2(x)$ for a certain fixed polynomial $p$,
where $f^2(x) =(f(x))^2$. This immediately implies that if $f\in\gapp$
then $\lambda x.2^{-p(|x|)}f^2(x)\in \sharpqp$. Their result can be
further refined to the following theorem.  This exemplifies a
characteristic feature of quantum gap functions.

\begin{theorem}\label{theorem:squared}{\rm (Squared Function Theorem)}\hs{1}
Assume that $K$ is admissible. If $f\in\gapqp_{K}$ then
$f^2\in\sharpqp_{K}$.
\end{theorem}

To prove Theorem \ref{theorem:squared}, we need the following key
lemma, called the Gap Squaring Lemma, which is compared to the
Squaring Lemma.

\begin{lemma}\label{lemma:Hadamard}{\rm (Gap Squaring Lemma)}\hs{2}
Assume that $K^*\subseteq K$. Let $M$ be any polynomial-time
synchronous dynamic normal-form unidirectional well-formed
$K$-amplitude QTM with a single final state. There exists another
polynomial-time synchronous dynamic normal-form unidirectional
well-formed $K$-amplitude QTM $N$ that, on any input $x$, halts in a
final superposition in which the amplitude of the configuration
$cf_{N,x}$ that consists of $x$ on the input tape, $1$ on the output
tape, and empty elsewhere is exactly $2\rho_M(x)-1$.
\end{lemma}

\begin{proof}
Let $M$ be the QTM given in the lemma. We define the desired QTM $N$
as follows.  Firstly, given input $x$, $N$ simulates $M$ on the same
input. Let $cf_{M,x}^{(0)}$ be the initial configuration of $M$ on
input $x$. Assume that $M$ halts in a final superposition
$\qubit{\phi} =
\sum_{y}\alpha_{x,y}\qubit{y}\qubit{b_y}$, where $y$ ranges over all
(valid) configurations (except for the content of the output tape) of
$M$ and the last qubit $\qubit{b_y}$ represents the content of $M$'s
output tape. The acceptance probability $\rho_{M}(x)$ thus equals
$\sum_{y:b_y=1}|\alpha_{x,y}|^2$.  Secondly, $N$ applies $-P_{\pi}$ to
$\qubit{b_y}$ and then we obtain the superposition $\qubit{\phi'}
=\sum_{y:b_y=1}\alpha_{x,y}\qubit{y}\qubit{1} -
\sum_{y:b_y=0}\alpha_{x,y}\qubit{y}\qubit{0}$.

By the Reversal Lemma, there exists a polynomial-time synchronous
dynamic normal-form unidirectional well-formed QTM $M_R$ that reverses
the computation of $M$. Note that $M_R$ also has $K$-amplitudes since
$K^*\subseteq K$.  Now, $N$ simulates $M_R$ starting with
$\qubit{\phi'}$ as its initial superposition. Note that if we run
$M_R$ on superposition $\qubit{\phi}$ then $M_R\qubit{\phi}$ becomes
the initial superposition $\qubit{cf_{M,x}^{(0)}}$. In our notation
$M_{R}\qubit{\phi}$, $M_R$ can be viewed as a unitary
operator. Abusing this notation, we write $M_{R}^{\dagger}$ to mean
the transposed conjugate of $M_R$. Observe that the inner product of
$\qubit{\phi}$ and $\qubit{\phi'}$ is $\measure{\phi}{\phi'} =
\sum_{y:b_y=1}|\alpha_{x,y}|^2 -
\sum_{y:b_y=0}|\alpha_{x,y}|^2$, which equals $2\rho_M(x)-1$.

Finally, $N$ outputs 1 (\ie acceptance) if it observes exactly
$\qubit{cf_{M,x}^{(0)}}$; otherwise, $N$ outputs 0 (\ie
rejection). The amplitude of the configuration $cf_{N,x}$ 
described in the
lemma is exactly $\bra{cf_{M,x}^{(0)}}M_{R}\ket{\phi'}$, which equals
$\bra{\phi}M_R^{\dagger}M_R\ket{\phi'}$. Since $N$ preserves the inner
product, we have $\bra{\phi}M_R^{\dagger}M_R\ket{\phi'} =
\measure{\phi}{\phi'} = 2\rho_M(x)-1$. This completes the proof.
\end{proof}

Now, we are ready to prove Theorem \ref{theorem:squared}.
\vs{2}

\begin{proofof}{Theorem \ref{theorem:squared}}
Let $f$ be any $\gapqp_{K}$-function. Using Lemma
\ref{lemma:conversion}, we obtain a polynomial-time synchronous
dynamic normal-form unidirectional well-formed $K$-amplitudes QTM $M$
with a single final state such that $f(x)= 2\rho_M(x)-1$ for all
$x$. Let $p$ be any polynomial satisfying that $M$ on input $x$ halts
at time $p(|x|)$. It follows from the Gap Squaring Lemma that there
exists another polynomial-time well-formed $K$-amplitude QTM $N$ that
starts on input $x$ and halts in a final superposition in which the
amplitude of configuration $cf_{N,x}$ is $2\rho_M(x)-1$, where
$cf_{N,x}$ is a unique configuration of $N$ that consists of $x$ on
the input tape, $1$ on the output tape, and empty elsewhere. Thus, the
acceptance probability of $N$ equals $(2\rho_{M}(x)-1)^2$, which is
obviously $f^2(x)$. Therefore, $f^2$ belongs to $\sharpqp_{K}$.
\end{proofof}

\subsection{Relationships between GapQP and GapP}

Quantum probability gap functions are closely related to their
classical counterpart.  The following theorem shows a differently
intertwined relationship between $\gapqp_{K}$ and $\gapp_{K}$
depending on the choice of amplitude set $K$.  If amplitudes are
restricted on rational numbers, then $\gapqp$ and $\gapp$ bear
fundamentally the same computational power. This indicates a
limitation of quantum computations.

Let $sign(a)$ be $0$, $1$, and $-1$ if $a=0$, $a>0$, and $a<0$,
respectively. Recall the identification of $\nat$ with $\Sigma^*$
given in \S\ref{sec:notions}.

\begin{theorem}\label{theorem:property}
\begin{enumerate}
\item For every $f\in\gapqp_{\complex}$, there exists a function
$g\in\gapp$ such that, for every $x$, $f(x)=0$ iff $g(x)=0$.
\vs{-2}
\item For every $f\in\gapqp_{\ptcomplex}$ and every polynomial $q$, 
there exist two functions $g\in\gapp$ and
$\ell\in\fp\cap\nat^{\Sigma^*}$ such that
$|f(x)-\frac{g(x)}{\ell(1^{|x|})}|\leq 2^{-q(|x|)}$ for all $x$.
\vs{-2}
\item For every $f\in\gapqp_{\algebraic}$, there exists a function 
$g\in\gapp$ such that $sign(f(x))= sign(g(x))$ for all $x$.
\vs{-2}
\item For every $f\in\gapqp_{\rational}$, there exist two functions 
$g\in\gapp$ and $\ell\in\fp\cap\nat^{\Sigma^*}$ such that
$f(x)=\frac{g(x)}{\ell(1^{|x|})}$ for all $x$.
\end{enumerate}
\end{theorem}

Theorem \ref{theorem:property}(4) generalizes a result of Fortnow and
Rogers \cite{FR99}, who considered only the amplitude set
$\{0,\pm1,\pm\frac{3}{5},\pm\frac{4}{5}\}$.

In what follows, we give the proof of Theorem \ref{theorem:property}.
Partly, we use the result by Yamakami and Yao \cite{YY99}, who
introduced a canonical representation of the amplitude $\alpha$ of each
configuration in a superposition of a QTM at time
$t$.  This representation makes it possible to encode such amplitude
into a finite sequence of integers and to simulate a quantum
computation by modifying such sequences in a classical manner.

Fix a well-formed QTM $M$ and let $D$ be the amplitude set of $M$. Let
$A=\{\alpha_1,\ldots,\alpha_m\}$ be the maximal subset of $D$ that is
algebraically independent. Define $F=\rational(A)$ and let $G$ be the
field generated by elements in $\{1\}\cup(D-A)$ over $F$. Let
$\{\beta_0,\beta_1,\ldots,\beta_{d-1}\}$ be a basis of $G$ over $F$
with $\beta_0=1$. Let $D'=D\cup\{\beta_i\beta_j\mid
i,j\in\integers_{d}\}$. Take any common denominator $u$ such that, for
every $\alpha\in D'$, $u\alpha$ is of the form
$\sum_{\svk}a_{\svk}(\prod_{i=1}^{m}\alpha_i^{k_i})\beta_{k_0}$, where
$\vk=(k_0,k_1,\ldots,k_m)$ ranges over $\integers_d\times
\integers^{m}$ and $a_{\svk}\in\integers$.  Thus, there exist two 
integers $e,d>0$ such that, the amplitude $\alpha$ of any
configuration $C$ in the superposition of $M$ at time $t$ on input
$x$, when multiplied by $u^{2t-1}$, can be of the form
$\sum_{\svk}a_{\svk}(\prod_{i=1}^{m}\alpha_i^{k_i})\beta_{k_0}$, where
$\vk=(k_0,k_1,\ldots,k_m)$ ranges over $\integers_{d}\times
(\integers_{[2et]})^{m}$ and $a_{\svk}\in\integers$. It is 
shown in \cite{YY99} that each $a_{\svk}$ is computed from $(x,\vk,C)$
by a certain $\gapp$-function $s$; namely, $s(x,\vk,C)=a_{\svk}$, and
thus, $\sum_{\svk}|a_{\svk}|\leq 2^{p(|x|,t)}$ for a certain fixed
polynomial $p$.

\ms

1) Let $f$ be any function in $\gapqp_{\complex}$. By the Squared
Function Theorem, $f^2$ belongs to $\sharpqp_{\complex}$. Consider the
set $A=\{x\mid f^2(x)>0\}$, which is in $\nqp_{\complex}$. As Yamakami
and Yao \cite{YY99} proved, $\nqp_{\complex}$-sets are all in
$\co\cequalp$. Thus, $A$ is also written as $A=\{x\mid g(x)\neq0\}$
for a certain $\gapp$-function $g$. Therefore, it immediately follows
that, for every $x$, $f(x)=0$ iff $g(x)=0$.

2) The essence of the following argument comes from \cite{ADH97}. We
begin with a key lemma.  For any QTM $M$ and any final configuration
$C$ of $M$ on input $x$, let $amp_{M}(x,C)$ denote the amplitude of
configuration $C$ in the final superposition of $M$ on input $x$ if
$M$ halts. The {\em complex conjugate} of $M$ is the QTM $M^*$ defined
exactly as $M$ except that its time-evolution operator $U_{M^*}$ is
the complex conjugate of $U_{M}$.

\begin{lemma}\label{lemma:amplitudes}
Assume that $K^*\subseteq K$.  Let $M$ be any polynomial-time
synchronous stationary well-formed $K$-amplitude QTM in normal form
with a single final state. There exists a well-formed QTM $N$ such
that, for every $x$, (1) $N$ halts in polynomial time with one symbol
from $\{0,1,\#\}$ written in the start cell of its output tape; (2)
$\sum_{C\in D_x^1}amp_{N}(x,C)=\rho_M(x)$; and (3) $\sum_{C\in
D_x^0}amp_{N}(x,C)=\overline{\rho}_M(x)$, where $D_x^i$ is the set of
all final configurations, of $N$ on $x$, whose output tape consists of
symbol $i\in\{0,1\}$ in the start cell.
\end{lemma}

\begin{proof}
The desired QTM $N$ works as follows. On input $x$, $N$ simulates $M$
on input $x$; when $M$ halts in a final configuration $C_1$, $N$
starts another round of simulation of $M^*$ in a different set of
tapes. After $M^*$ reaches a final configuration $C_2$, $N$
deterministically checks if both final configurations $C_1$ and $C_2$
are identical. If $C_1\neq C_2$, then $N$ outputs the blank symbol \#
and halts.  Now, assume that $C_1=C_2$. If this unique configuration
$C_1$ is an accepting configuration, then $N$ outputs $1$; otherwise,
it outputs $0$. On this computation path, we obtain the amplitude
$amp_{M}(x,C_1)\cdot amp_{M^*}(x,C_2)$, which equals
$|amp_{M}(x,C_1)|^2$. For each $i\in\{0,1\}$, let $E^i_x$ denote the
set of all final configurations, of $M$ on $x$, in which the output
tape consists of symbol $i$ in the start cell. Thus, the sum
$\sum_{C\in D^1_x}amp_{N}(x,C)$ equals $\sum_{C\in
E^1_x}|amp_{M}(x,C)|^2$, which is exactly $\rho_{M}(x)$. Similarly,
$\sum_{C\in D^0_x}amp_{N}(x,C)$ equals $\overline{\rho}_{M}(x)$.
\end{proof}

Let $f\in\gapqp_{\ptcomplex}$ and take a polynomial-time well-formed
$\ptcomplex$-amplitude QTM $M$ that witnesses $f$. We can assume from
Lemma \ref{lemma:conversion} that $M$ is further synchronous,
stationary, and in normal form. Let $q$ be any polynomial. 
By Lemma \ref{lemma:amplitudes}, there exists a polynomial-time
well-formed QTM $M$ with $\ptcomplex$-amplitudes such that $f(x)$
equals $\sum_{C\in D_x^1}amp_{M}(x,C) -\sum_{C\in D_x^0}amp_{M}(x,C)$,
where each $D_x^i$ is defined in Lemma \ref{lemma:amplitudes}.  Let
$r$ be any polynomial that bounds the running time of $M$. Assume that
$|D_x^0\cup D_x^1|\leq 2^{r(|x|)}$ and $r(|x|)\geq \log{r(|x|)}+|x|$
for any $x$.

Let $n$ be any sufficiently large integer, $x$ be any string of length
$n$, and $C$ be any final configuration of $M$ on input $x$.  Let
$\ell(x)=2^{2r(n)+q(n)+1}$ and $h(x,C)=amp_{M}(x,C)$.  Assume first
that there exists a function $\tilde{h}$ in $\gapp$ such that
$|h(x,C)-\frac{\tilde{h}(x,C)}{\ell(1^n)}|\leq 2^{-r(n)-q(n)-1}$,
which implies $|\sum_{C\in D_x^i}g(x,C)-
\sum_{C\in D_x^i}\frac{\tilde{h}(x,C)}{\ell(1^n)}|
\leq 2^{r(n)}\cdot 2^{-r(n)-q(n)-1} = 2^{-q(n)-1}$ for each 
$i\in\{0,1\}$. The desired $\gapp$-function $g$ is then defined as
$g(x)=\sum_{C\in D_x^1}\tilde{h}(x,C)-\sum_{C\in
D_x^0}\tilde{h}(x,C)$. It follows that $|f(x)-\frac{g(x)}{\ell(1^n)}|
\leq |\sum_{C\in D^1_x}h(x,C) - \frac{\tilde{h}(x,C)}{\ell(1^n)}| + 
|\sum_{C\in D^0_x}h(x,C) - \frac{\tilde{h}(x,C)}{\ell(1^n)}|
 \leq 2^{-q(n)}$, as requested.

To complete the proof, we show the existence of $\tilde{h}$.  Recall
that amplitude $amp_{M}(x,C)$, when multiplied with $u^{2r(n)-1}$, is
of the form
$\sum_{\svk}a_{\svk}(\prod_{i=1}^{m}\alpha_i^{k_i})\beta_{k_0}$, where
$\vk=(k_0,\ldots,k_m)$ is taken over $\integers_{d}\times
(\integers_{[2er(n)]})^m$ and $a_{\svk}\in\integers$. Note that the
number of such $\vk$'s is $d(4er(n))^m$, which is at most
$2^{r(n)}$. Note also that the complex numbers
$\alpha_1,\ldots,\alpha_m,\beta_0,\ldots,\beta_{d-1},u$ are all in
$\ptcomplex$. Thus, these numbers can be approximated by
certain polynomial-time deterministic TMs with any desired
precision. By simulating such machines in polynomial time, we can
compute an approximation $\tilde{\rho}_{x,\svk,C}$ of the value
$(\prod_{i=1}^{m}\alpha_i^{k_i})\beta_{k_0}u^{1-2r(n)}$ to within
$2^{-2r(n)-q(n)-1}$. Let $j(x,\vk,C)$ be the integer closest to
$\ell(1^n)a_{\svk}\tilde{\rho}_{x,\svk,C}$. The function
$\tilde{h}(x,C)$ defined as $\sum_{\svk}j(x,\vk,C)$ satisfies
$|h(x,C)-\frac{\tilde{h}(x,C)}{\ell(1^n)}|\leq
\sum_{\svk}2^{-2r(n)-q(n)-1}\leq 2^{-r(n)-q(n)-1}$. By its definition,
$\tilde{h}$ belongs to $\gapp$.

3) Let $f\in\gapqp_{\algebraic}$. By Theorem
\ref{theorem:property}(1), there exists a function $g_0\in\gapp$ such
that, for every $x$, $g_0(x)=0$ iff $f(x)=0$.  Let $M$ be any
$\algebraic$-amplitude well-formed QTM that witnesses $f$ in time
polynomial $p$. Let $x$ be any input of length $n$. Since $M$ has
$\algebraic$-amplitudes, the amplitude $\alpha$ of each configuration
in the final superposition of $M$ at time $p(n)$ on input $x$, when
multiplied by $u^{2p(n)-1}$, has the form
$\sum_{\svk}a_{\svk}(\sum_{i=0}^{m}\alpha_i^{k_i})$, where
$\vk=(k_0,k_1,\ldots,k_m)$ ranges over $\integers_{d}\times
(\integers_{[2ep(n)]})^{m}$ and $a_{\svk}\in\integers$.

We use the following lemma on an approximation of a polynomial of
algebraic numbers.

\begin{lemma}\label{lemma:algebra}{\rm (cf. \cite{Sto74})}\hs{1}
Let $\alpha_1,\ldots,\alpha_m\in\algebraic$. Let $d$ be the degree of
$\rational(\alpha_1,\ldots,\alpha_m)/\rational$. There exists a
constant $c>0$ that satisfies the following for any complex number
$\alpha$ of the form
$\sum_{\svk}a_{\svk}(\prod_{i=1}^{m}\alpha_i^{k_i})$, where
$\vk=(k_1,\ldots,k_m)$ ranges over $\integers_{[N_1]}\times
\cdots\times \integers_{[N_m]}$, $(N_1,\ldots,N_m)\in\nat^m$, and
$a_{\svk}\in\integers$. If $\alpha\neq0$ then $|\alpha|\geq
(\sum_{\svk}|a_{\svk}|)^{1-d}\prod_{i=1}^mc^{-d N_i}$.
\end{lemma}

By Lemma \ref{lemma:algebra}, any nonzero amplitude $\alpha$ of a
configuration in the final superposition of $M$ on $x$ has the squared
magnitude $\geq\frac{1}{|u|^{2p(n)-1}}
(\sum_{\svk}|a_{\svk}|)^{1-d}\prod_{i=1}^{m}c^{-2edp(n)}$. Note that
$\sum_{\svk}|a_{\svk}|$ and $|u|^{2p(n)-1}$ are both bounded above by
an exponential in $n$. This yields a lower bound of the absolute value
$|f(x)|$ when $f(x)\neq0$. By choosing an appropriate polynomial $s$,
we thus obtain $|f(x)|\geq 2^{-s(|x|)}$ for all $x$.

By Theorem \ref{theorem:property}(2), there are two functions
$k\in\gapp$ and $\ell\in\fp\cap\nat^{\Sigma^*}$ satisfying that
$|f(x)-\frac{k(x)}{\ell(1^{|x|})}|\leq 2^{-s(|x|)-1}$ for all $x$.
Consider the case where $f(x)>0$. Since $f(x)\geq 2^{-s(n)}$, it
follows that $\frac{k(x)}{\ell(1^n)}\geq 2^{-s(n)-1}>0$. In the case
where $f(x)<0$, since $f(x)\leq -2^{-s(n)}$, we have
$\frac{k(x)}{\ell(1^n)}\leq -2^{-s(n)-1}<0$. Therefore, $f(x)>0$
implies $k(x)>0$ and $f(x)<0$ implies $k(x)<0$.

Finally, the desired function $g$ is defined by $g(x)=g_0(x)^2\cdot
k(x)$ for all $x$. Obviously, $g$ is in $\gapp$ since $g_0$ and $k$
are both $\gapp$-functions.

4) In the proof of Theorem \ref{theorem:property}(2), if in addition
$M$ has $\rational$-amplitudes, then the value
$a_{\svk}(\prod_{i=1}^{m}\alpha_i^{k_i})\beta_{k_0}u^{1-2r(n)}$, when
multiplied with $\ell(1^n)$, becomes an integer and thus, we can
precisely compute it in polynomial time.  Therefore, $\tilde{h}$
satisfies that $h(x,C)=\frac{\tilde{h}(x,C)}{\ell(1^n)}$, and
consequently, $f(x)=\frac{g(x)}{\ell(1^n)}$.

\vs{1}

This completes the proof of Theorem \ref{theorem:property}.

\section{Quantum Functions with an Access to Oracles} 

An {\em oracle} is in general an external device that provides an
underlying computation with extra information by means of oracle
queries. The role of oracles in quantum computation was recognized as
far back as the early 1990s by Deutsch and Jozsa \cite{DJ92}. Many
existing quantum algorithms in essence use oracle queries in order to
access inputs and the number of oracle queries is used to measure the
complexity of these quantum algorithms. This section introduces {\em
relativized quantum functions} that can access oracles in two
different manners: adaptive and nonadaptive queries.

\subsection{Adaptive Queries and Nonadaptive Queries}

We first give a general resource-bounded query model for relativized
quantum functions. For a later use, a restriction of the number of
queries is imposed on every computation path of a given oracle
QTM. {}From such a restriction arises the notion of {\em bounded
queries}.

In what follows, $r$ denotes an arbitrary function in $\nat^{\nat}$
and $R$ is any subset of $\nat^{\nat}$. In this paper, an {\em oracle}
means a subset of $\Sigma^*$ and $\CC$ denotes an arbitrary class of
oracles.

\begin{definition}\label{def:adaptive}
Let $A$ be any oracle. A function $f$ is in $\feqp^{A[r]}$ if there
exists a polynomial-time well-formed oracle QTM $M$ such that, for
every $x$, $M$ on input $x$ outputs $f(x)$ with certainty using oracle
$A$ and makes at most $r(|x|)$ queries on each computation path. Let
$\feqp^{\CC[r]}$ be the union of $\feqp^{A[r]}$'s for all $A\in
\CC$. The class $\feqp^{A[R]}$ ($\feqp^{\CC[R]}$, resp.) is the union
of $\feqp^{A[r]}$'s ($\feqp^{\CC[r]}$'s, resp.) for all $r\in
R$. Conventionally, when $R=\nat^{\nat}$, we write $\feqp^{A}$
($\feqp^{\CC}$, resp.) instead of $\feqp^{A[R]}$ ($\feqp^{\CC[R]}$,
resp.).  Similar notions are introduced to $\fbqp$, $\sharpqp$, and
$\gapqp$.
\end{definition}

The oracle QTM $M$ in Definition \ref{def:adaptive} 
is said to make {\em
adaptive} (or {\em sequential}) {\em queries} since the choice of a
query word relies on the oracle answers to its previous queries. By
contrast, we can define an oracle QTM that makes {\em
nonadaptive} (or {\em parallel}) {\em queries} where all query words
are pre-determined before the first oracle query. Our nonadaptive
query model\footnote{The nonadaptive query model was independently
introduced in \cite{BD99}.} is an immediate adaptation of
$\np_{\|}^{A}$ (see, \eg \cite{Wag90}). Every computation path $P$
generates on a designated tape a {\em query list}---a list of all
query words (separated by a special separator) that are possibly
queried along computation path $P$ before any query 
is made on this path.

There are three important issues concerning the definition of parallel
queries in a quantum setting. The first issue is the timing of the
completion of all query lists. Quantum interference makes it
possible for two different computation paths to interfere. 
Destructive interference in particular annihilates certain
configurations. Hence, we need to 
avoid the case where the first query is made
at a computation path $P_1$ but a query list on another computation
path $P_2$ is not yet finished because any query list on path
$P_2$ may be affected by the result of the queries made earlier on
path $P_1$ due to quantum interference. An important requirement of
parallel queries is that an oracle 
QTM should complete all query lists just
before it enters the pre-query state {\em for the first time} in its
entire computation tree. At this moment, we say that all the query
lists are {\em completed}. Once the query list on each computation
path is completed, $M$ can freely delete any word from this list but
cannot add any word to the list afterward.

The second issue concerns the ``actual'' queries 
compared to the query words
generated in a query list. In a classical setting, we can always
assume that all query words in a query list are indeed queried
whether or not we use their oracle answers. 
Nonetheless, the oracle QTM may
not properly perform quantum interference if the machine keeps
unnecessary oracle answers on its tapes.  Thus, the classical
requirement would be relaxed so that all the query words in each query
list are not necessarily queried during a computation.

The last issue is the maintenance of query lists since maintaining a
query list until the end of the computation may prohibit any quantum
interference to occur 
during a computation that follows.  The completed query
list on any computation path $P$ is 
allowed to alter after the first query
is made along path $P$ in order to make this path interfere with other
computation paths that had produced different query lists.

\begin{definition}
The class $\feqp_{\|}^{A[r]}$ is the subset of $\feqp^{A[r]}$ with the
extra condition that, on each computation path, just before $M$ enters
a pre-query state for the first time in the entire computation, it
completes all query lists. Any query list completed on each
computation path must be maintained unaltered until the first query 
is made on
this computation path but the list 
may be altered once the machine makes the
first query on this computation path. All the words in the query list
may not be queried but any word that is queried must be in the query
list. The class $\feqp_{\|}^{\CC[r]}$ is the union of
$\feqp_{\|}^{A[r]}$'s over all $A\in \CC$. The notation
$\feqp_{\|}^{A[R]}$ ($\feqp_{\|}^{\CC[R]}$, resp.) denotes the union
of $\feqp_{\|}^{A[r]}$'s ($\feqp_{\|}^{\CC[r]}$, resp.) over all $r\in
R$. Similar notions are introduced to $\fbqp$, $\sharpqp$, and
$\gapqp$.
\end{definition}

The following lemma is immediate.

\begin{lemma}\label{lemma:query-FEQP}
1. $\feqp = \fp_{\|}^{\eqp} = \fp^{\eqp} = \feqp_{\|}^{\eqp} =
\feqp^{\eqp}$.

2. $\fbqp = \fp_{\|}^{\bqp} = \fp^{\bqp} = \fbqp_{\|}^{\bqp} =
\fbqp^{\bqp}$.

3. $\qmasv \subseteq \fp_{\|}^{\qma}$.

4. $\sharpqp = \sharpqp_{\|}^{\eqp} = \sharpqp^{\eqp}$.

5. $\gapqp = \gapqp_{\|}^{\eqp} = \gapqp^{\eqp}$.
\end{lemma}

\begin{proof}
1) It easily follows that $\fp_{\|}^{\eqp}\subseteq \fp^{\eqp}\cup
\feqp_{\|}^{\eqp} \subseteq \feqp^{\eqp}$. To show that
$\feqp\subseteq \fp_{\|}^{\eqp}$, let $f\in\feqp$ and let $p$ be any
polynomial such that $|f(x)|\leq p(|x|)$ for all $x$. Define
$A=\{\pair{x,1^i}\mid
\mbox{the $i$th bit of $f(x)$ is 1}\}$ and $B=\{\pair{x,1^j}\mid
|f(x)|\geq j \}$. The last set $B$ is necessary to determine the
length of $f(x)$.  We can show that $A$ and $B$ are both in $\eqp$ by
simulating the QTM that computes $f$. Thus, $A\oplus B\in\eqp$.  Now,
it is easy to show that $f\in\fp_{\|}^{A\oplus B}$ by making
nonadaptive queries
$\pair{x,1},\pair{x,11},\ldots,\pair{x,1^{p(|x|)}}$ to both $A$ and
$B$.

It still remains to prove that $\feqp^{\eqp}\subseteq\feqp$. Let
$f\in\feqp^A$ for a certain oracle $A$ in $\eqp$. Let $M$ be any
polynomial-time well-formed oracle QTM that, on input $x$, outputs
$f(x)$ with certainty. The Canonical Form Lemma allows $M$ to be in a
canonical form with oracle $A'$. Since $A'\in\eqp$, by Lemma
\ref{lemma:conversion}, $A'$ is recognized with probability $1$ 
by a certain polynomial-time synchronous
dynamic stationary normal-form unidirectional well-formed
$\ptcomplex$-amplitude QTM $N$ with a single final state.
We further assume from the Squaring
Lemma that $M$'s final superposition consists entirely of a
configuration, with amplitude $1$, in which $M$ is in a final state,
$M$'s output tape holds only one bit in the start cell, and all other
tapes are empty. Such a configuration can be identified with a bit
written on the output tape. Consider the quantum algorithm $\QQ$ that
simulates $M$ on input $x$ and, whenever it invokes a query $y$,
simulates $N$ on input $y$. This algorithm $\QQ$ can be implemented on
a certain well-formed oracle QTM since $M$ makes the same number of
queries to oracle $A$ with query words of the same length along each
computation path on any input of fixed length. This implies that
$f\in\feqp$.

2) Similar to 1) except for the proof of $\fbqp^{\bqp}=\fbqp$. We can
show $\fbqp^{\bqp}=\fbqp$ in a way similar to $\bqp^{\bqp}=\bqp$
\cite{BBBV97} by amplifying the success probability of a QTM, which
computes a given oracle set, from $3/4$ to close to $1$ so that the
cumulative error is still bounded above by $1/4$ after the
polynomially-many runs of this QTM.

3) Let $f\in\qmasv$, which is witnessed by a certain polynomial $p$
and a polynomial-time well-formed QTM $M$ as in Definition
\ref{def:QMASV}. Let $q$ be any polynomial satisfying $|f(x)|\leq
q(|x|)$ for all $x$. We modify the definitions of $A$ and $B$ in 1) as
follows. Let $A$ be the collection of all strings $\pair{x,1^i}$,
where $x\in\Sigma^*$ and $0\leq i\leq q(|x|)$, such that there exist a
string $y\in\Sigma^{q(|x|)}$ and a qustring
$\qubit{\phi}\in\Phi_{p(|x|)}$ satisfying that $M$ on input
$\qubit{x}\qubit{\phi}$ outputs $\qubit{1}\qubit{y}$ with probability
at least $3/4$ with the additional condition that the $i$th bit of $y$
must be $1$.  The set $B$ is defined similar to $A$ but it checks if
$M$ on input $\qubit{x}\qubit{\phi}$ outputs $\qubit{1}\qubit{y}$ with
$|y|\geq i$ with probability at least $3/4$.  It is easy to see that
$A$ and $B$ are in $\qma$ because of the choice of $M$. Similar to 1),
making appropriate nonadaptive queries to $A\oplus B$ 
computes $f(x)$ in polynomial time.

4) and 5) These proofs are similar to 1).
\end{proof}

As an immediate consequence of Lemma \ref{lemma:query-FEQP}, we can
characterize $\eqp$ as the collection of all low sets for $\sharpqp$
or for $\gapqp$. This contrasts the classical results
$\low\sharpp=\up\cap\co\up \subseteq \spp=\low\gapp$ \cite{FFK94}.

\begin{corollary}
$\eqp = \low\sharpqp_{\|} = \low\sharpqp = \low\gapqp_{\|} = \low\gapqp$.
\end{corollary}

\begin{proof}
Clearly, $\low\gapqp\subseteq \low\gapqp_{\|}$. Since
$\gapqp^{\eqp}=\gapqp$ by Lemma \ref{lemma:query-FEQP}(5), it follows
that $\eqp\subseteq \low\gapqp$. We still need to prove that
$\low\gapqp_{\|}\subseteq\eqp$. Let $A$ be any set in
$\low\gapqp_{\|}$. It is easy to see that
$\chi_{A}\in\gapqp_{\|}^{A[1]}$. Since $\gapqp_{\|}^A\subseteq\gapqp$,
we obtain that $\chi_{A}\in\gapqp$.  By the Squared Function Theorem,
$\chi_A^2$ is in $\sharpqp$. Since $\chi_A^2=\chi_A$, $\chi_{A}$ also
belongs to $\sharpqp$. This yields the desired conclusion that
$A\in\eqp$.  Therefore, $\low\gapqp_{\|}\subseteq\eqp$. Similarly, we
can show that $\eqp=
\low\sharpqp_{\|} = \low\sharpqp$.  
\end{proof}

A wide gap has been exhibited between a function class and a language
class in a classical setting; for instance,
$\p_{\|}^{\np}=\p^{\np[O(\log )]}$ \cite{Wag90} but
$\fp_{\|}^{\np}\neq\fp^{\np[O(\log n)]}$ if $\np\neq\rp$
\cite{JT95}. Quantum interference, on the contrary, draws such two
classes close together. The following proposition is an adaptation of
the argument in \cite{BD99}, in which an quantum algorithm of
Bernstein and Vazirani \cite{BV97} is effectively used.

\begin{proposition}
Let $R\subseteq \{0,1\}^{\Sigma^*}$ and assume that $R$ is closed
under constant multiplication. For any oracle $A$,
$\fbqp_{\|}^A\subseteq \fbqp^{A[R]}$ iff $\bqp_{\|}^A\subseteq
\bqp^{A[R]}$.
\end{proposition}

\begin{proof}
The implication from left to right is obvious. Let $f$ be any function
in $\fbqp_{\|}^A$. Assuming that $\bqp_{\|}^A\subseteq \bqp^{A[R]}$,
we want to show that $f$ belongs to $\fbqp^{A[R]}$. Let $p$ be any
polynomial that bounds the length of the value of $f$. Without loss of
generality, we assume that $f$ is length-regular since, otherwise,
we can set $\tilde{f}(x)=f(x)10^{p(n)-|f(x)|}$ for all $x$. For
simplicity, assume that $|f(x)|=p(|x|)$ for all $x$.

Define $B=\{\pair{x,z} \mid b\in\{0,1\},|z|=|f(x)|,f(x)\cdot z=1\}$,
where $u\cdot v$ is the {\em dot product} of $u$ and $v$. It follows
from $f\in\fbqp_{\|}^A$ that $B$ is in $\bqp_{\|}^A$. By our
assumption, $B$ is also in $\bqp^{A[r]}$ for a certain function $r\in
R$. Since Lemma \ref{lemma:conversion} relativizes, there exists a
polynomial-time synchronous dynamic stationary normal-form
unidirectional well-formed oracle QTM $M_0$, with a single final
state, that recognizes $B$ with oracle $A$ with error probability
$\leq 1/4$. We first amplify its success probability from $3/4$ to
$\sqrt{79/80}$. For such a QTM, we apply the Squaring Lemma (for an
oracle QTM) and obtain another QTM $M_1$. We modify this $M_1$ so
that, on input $\qubit{x}\qubit{z}\qubit{b}$, it produces a final
superposition of configurations, one of which has only
$\qubit{x}\qubit{z}\qubit{b\oplus \chi_{B}(\pair{x,z})}$ written on
the tapes with positive real amplitude $\geq \sqrt{79/80}$. Obviously,
$M_1$ makes only $O(r(n))$ queries.

The new QTM $N$ works as follows.  On input $x$ of length $n$, write
$\qubit{0^{p(n)}}\qubit{1}$ on a new blank tape and apply $H^{\otimes
p(n)}\otimes H$. We then have
$2^{-p(n)/2}\sum_{z:|z|=p(n)}\qubit{z}\otimes\qubit{\phi^{-}}$, where
$\qubit{\phi^{-}}=\frac{1}{\sqrt{2}}(\qubit{0}-\qubit{1})$. For each
$\qubit{z}\qubit{b}$, where $b\in\{0,1\}$, run $M_1^A$ to change
$\qubit{x}\qubit{z}\qubit{b}$ to
$\sqrt{79/80}\qubit{x}\qubit{z}\qubit{b\oplus (f(x)\cdot z)}
+\qubit{\psi_{x,z,b}}$, where $\qubit{\psi_{x,z,b}}$ is a certain
qustring.  At this moment, we obtain
$2^{-p(n)/2}[\sum_{z}(-1)^{f(x)\cdot
z}\sqrt{79/80}\qubit{x}\qubit{z}\qubit{\phi^{-}}+
\sum_{z,b}\qubit{\psi_{x,z,b}}]$. Apply $I^{\otimes n}\otimes
H^{\otimes p(n)}\otimes I$. The final superposition becomes
$\sqrt{79/80}\qubit{x}\qubit{f(x)}\qubit{\phi^{-}}+\qubit{\psi'}$ for
a certain qustring $\qubit{\psi'}$ since $H^{\otimes
p(n)}(2^{-p(n)/2}\sum_{z}(-1)^{f(x)\cdot z}\qubit{z}) =
\qubit{f(x)}$. Unfortunately, $\qubit{\psi'}$ is not known to be
orthogonal to $\qubit{x}\qubit{f(x)}\qubit{\phi^{-}}$. However, since
$\|\qubit{\psi'}\|\leq \sqrt{1/80}$, we can observe
$\qubit{x}\qubit{f(x)}$ with probability at least
$(\sqrt{79/80}-\sqrt{1/80})^2\geq 3/4$.  Thus,
$f\in\fbqp^{A[O(r(n))]}\subseteq
\fbqp^{A[R]}$.
\end{proof}

\subsection{Oracle Separation}

Relativizations of complexity classes have become substantial topics
in quantum complexity theory
\cite{BBBV97,BV97,BB94,FR99,GP01,Sim97,Wat00}.  Berthiaume and
Brassard \cite{BB94} in particular constructed an oracle $A$ such that
$\p^A\neq \eqp_{\|}^{A[1]}$ using the quantum algorithm of Deutsch and
Jozsa \cite{DJ92}. By refining their result, we show the existence of
a set $A$ such that $\feqp_{\|}^{A[1]}\nsubseteq \sharpe^A$, which
immediately implies $\feqp^{A[1]}_{\|}\nsubseteq \fp^A$ since
$\fp^A\subseteq\sharpe^A$.

\begin{proposition}\label{prop:separation}
There exists an oracle $A$ such that
$\feqp_{\|}^{A[1]}\cap\{0,1\}^{\Sigma^*} \nsubseteq\sharpe^A$.
\end{proposition}

\begin{proof}
We say that a set $A$ is {\em good} if, for every $n\in\nat$, either
$|A\cap\Sigma^{n^2}|=|\Sigma^{n^2}
\setminus A|$ or $|A\cap\Sigma^{n^2}|\cdot |\Sigma^{n^2} \setminus
A|=0$.  For any set $A$ and any string $x$, let
$f^A(x)=2^{-2|x|^2}\cdot(|A\cap\Sigma^{|x|^2}| - |\Sigma^{|x|^2}
\setminus A|)^2$.  To compute this function $f^A$, consider the
following oracle QTM $N$ with oracle $A$.
\begin{quote}
{\sf On input $x$ of length $n$, write $\qubit{0^{n^2}}\qubit{1}$ on a
query tape and apply $H^{\otimes n^2}\otimes H$. Copy the first $n^2$
bits into a query list on a designated tape. 
Invoke an oracle query.  Delete the query
list. Again, apply $H^{\otimes n^2}\otimes I$. Observe the first $n^2$
bits on the query tape. Output $1$ if $\qubit{0^{n^2}}$ is observed, and 
output $0$ otherwise.}
\end{quote}
The deletion of each query list is possible since the query list
contains the exact copy of the first $n^2$ bits on the query tape. It
follows by a simple calculation that $N^A$ on input $x$ outputs
$f^{A}(x)$ with certainty if $A$ is good.  Thus, $f^A$ belongs to
$\feqp_{\|}^{A[1]}\cap\{0,1\}^{\Sigma^*}$ for any good oracle $A$.
 
Subsequently, we construct a good oracle $A$ such that
$f^A\not\in\sharpe^A$. For our construction, we need an {\em
effective} enumeration of all $2^{O(n)}$-time bounded nondeterministic
TMs. Let $\{M_i\}_{i\in\nat}$ be such an enumeration and define
$\{c_i\}_{i\in\nat}$ to be an enumeration of natural numbers (with
possible repetition) such that each 
$M_i$ halts within time $2^{c_i(n+1)}$ on
all inputs of length $n$, independent of the choice of oracles.  We
construct the desired oracle $A$ stage by stage.

Initially, set $n_{-1}=0$ and $A_{-1}=\setempty$. At stage $i\in\nat$
of the construction of $A$, let $n_i$ denote the minimal integer
satisfying that $2^{c_{i-1}(n_{i-1}+1)} <n_i$ and $c_i(n_i+1)<n_i^2
-1$. Assuming $A_{i-1}\subseteq\Sigma^{\leq n_{i-1}^2}$, we define $B
=A_{i-1} \cup
\Sigma^{n_i^2}$. Clearly, $f^B(0^{n_i})=1$. If
$\#M_i^{B}(0^{n_i})\neq1$, then define $A_i$ to be $B$. Assume
otherwise. There exists a unique accepting computation path $P$ of
$M_i$ on $0^{n_i}$. Let $Q_{P}$ denote the set of all words that $M_i$
queries along this computation path $P$. Since $|Q_{P}|\leq
2^{c_i(n_i+1)}<2^{n_i^2-1}$, there is a subset $C$ of $\Sigma^{n_i^2}$
such that $Q_{P}\cap\Sigma^{n_i^2}\subseteq C$ and $|C\cap
\Sigma^{n_i^2}|=|\Sigma^{n_i^2}\setminus C|$. For this $C$,
$\#M_i^{A_{i-1}\cup C}(0^{n_i})\geq1$ but $f^{A_{i-1}\cup
C}(0^{n_i})=0$. Thus, we should set $A_i=C$. After all the stages,
define $A=\bigcup_{i\in\nat}A_i$. This set $A$ satisfies the
proposition.  
\end{proof}

Proposition \ref{prop:separation} demonstrates a strength of the
nonadaptive query class $\feqp_{\|}^{A[1]}$ over the adaptive query
class $\sharpe^A$. On the contrary, we show a limitation of
$\sharpqp_{\|}^A$ by exhibiting the existence of an oracle $A$ that
makes $\fp^{A[n]}$ more powerful than $\sharpqp_{\|}^A$, where
$\fp^{A[n]}$ is an abbreviation of $\fp^{A[\lambda n.n]}$.

\begin{theorem}\label{theorem:separation-FP}
There exists a set $A$ such that
$\fp^{A[n]}\cap\{0,1\}^{\Sigma^*} \not\subseteq\sharpqp_{\|}^A$.
\end{theorem}

\begin{proof}
We begin with the definition of a test function $f$. For each string
$z\in\Sigma^{\geq3}$, let $z_A =
\linebreak
\chi_A(z0^{|z|-2})\chi_A(z0^{|z|-3}1)\chi_A(z0^{|z|-4}11)\cdots
\chi_A(z1^{|z|-2})$. Note that $|z_A|=|z|-1$. For completeness, 
whenever $z\in\Sigma^{\leq2}$, set $z_A=\lambda$. The desired function
$f^A$ is defined as $f^A(x)=
\chi_A(x0^{|x|}_A)$ for each $x\in\Sigma^*$ and $A\subseteq\Sigma^*$. 
Since $f^A(x)\in\{0,1\}$ for all $x$ and $A$,
$f^A$ is in $\fp^{A[n]}\cap\{0,1\}^{\Sigma^*}$.

To complete the proof, it suffices to construct a set $A$ satisfying
that $f^A\not\in\sharpqp_{\|}^A$. Let $\{M_i\}_{i\in\nat}$ and
$\{p_i\}_{i\in\nat}$ be respectively two effective enumerations of all
polynomial-time well-formed oracle QTMs and of all polynomials such
that each $M_i$ halts within time $p_i(n)$ on all inputs of length $n$
independent of the choice of oracles. We build by stage a series of
disjoint sets $\{A_i\}_{i\in\nat}$ and then define
$A=\bigcup_{i\in\nat}A_i$. This $A$ satisfies the theorem.

For convenience, set $n_{-1}=3$ and $A_{-1}=\setempty$. Consider stage
$i\in\nat$. Let $n_i$ be the minimal integer such that
$p_{i-1}(n_{i-1}) <n_i$ and $8p_i(n_i)^4<2^{n_i}$. In the case where
$M_i$ does not make valid nonadaptive queries to a certain oracle
$A\cup A_{i-1}$ with $A\subseteq \Sigma^{2n_i-2}\cup\Sigma^{2n_i-1}$,
we set $A_i$ as this $A$ and go to the next stage. Hereafter, we
assume that $M_i$ makes nonadaptive queries to any oracle of the form
$A\cup A_{i-1}$ with $A\subseteq \Sigma^{2n_i-2}\cup\Sigma^{2n_i-1}$.
Now, we want to show the existence of a set
$A\subseteq\Sigma^{2n_i-2}\cup\Sigma^{2n_i-1}$ such that
$\chi_{A}(0^{n_i}0^{n_i}_{A})\neq
\rho^{A\cup A_{i-1}}_{M_i}(0^{n_i})$. Assume otherwise that 
$\chi_{A}(0^{n_i}0^{n_i}_A)=\rho^{A\cup A_{i-1}}_{M_i}(0^{n_i})$ for
any set $A\subseteq\Sigma^{2n_i-2}\cup\Sigma^{2n_i-1}$, and draw a
contradiction.  For readability, we omit subscript $i$ in the
following argument.

Let $S$ be the set of all strings $y\in\Sigma^{n-1}$ such that at
least one of the query lists of $M$ on input $0^n$ include word
$0^ny$. Note that $S$ does not depend on the choice of oracles since
$M$ makes nonadaptive queries to any oracles of the form $A\cup
A_{i-1}$ with $A\subseteq\Sigma^{2n-2}\cup\Sigma^{2n-1}$. We first
claim that $|S|=2^{n-1}$ since, otherwise, we can choose an
appropriate oracle $A$ such that $\chi_{A}(0^n0^n_A)\neq \rho^{A\cup
A_{i-1}}_M(0^n)$.

For each $y\in S$, let $\tilde{q}_y$ be the sum of all squared
magnitudes of $M$'s configurations $cf$ in any superposition of $M$ on
input $0^n$ where $cf$ has a query list containing word $0^ny$.  Note
that each query list consists of at most $p(n)$ words.  It thus
follows that $\sum_{y\in S}\tilde{q}_y \leq
p(n)\sum_{j=1}^{p(n)-1}\|\qubit{\phi_j}\|^2 \leq p(n)^2$, where
$\qubit{\phi_j}$ is the superposition of $M$'s configurations at time
$j$ on input $0^n$.  Recall that $q^t_z(M,A,u)$ is the query magnitude
of string $z$ of $M^A$ on input $u$ at time $t$. Let $y$ be any string
in $S$ and fix $A$ that satisfies $y=0^n_A$. Moreover, let $A_y$ be
$A$ except that $\chi_{A_y}(0^ny)=1-\chi_{A}(0^ny)$. Note that
$A\triangle A_y=\{0^ny\}$. It follows by our assumption that
$\rho^{A\cup A_{i-1}}_{M}(0^{n})= 1- \rho^{A_y\cup
A_{i-1}}_{M}(0^{n})$.  By Lemma \ref{lemma:query-magnitude}, since
$|\rho^{A\cup A_{i-1}}_{M}(0^{n}) -
\rho^{A_y\cup A_{i-1}}_{M}(0^{n})|=1$, 
we have $\sum_{j=1}^{p(n)-1} q^j_{0^ny}(M,A\cup A_{i-1},0^n)
\geq\frac{1}{4p(n)}$. Clearly, $q^j_{0^ny}(M,A\cup A_{i-1},0^n) \leq
\tilde{q}_y$ for each $j$ since $M$ makes nonadaptive queries. Thus,
$\sum_{j=1}^{p(n)-1}q^j_{0^ny}(M,A\cup A_{i-1},0^n) \leq
p(n)\tilde{q}_y$, which implies $\tilde{q}_y\geq
\frac{1}{4p(n)^2}$.  This immediately draws the conclusion that
$|S|\leq 4p(n)^4$ since $|S|\cdot\min_{y\in S}\{\tilde{q}_y\}\leq
\sum_{y\in S}\tilde{q}_y$. This contradicts the fact $|S|=2^{n-1}$
since $8p(n)^4<2^n$.  
\end{proof}

{}From Proposition
\ref{prop:separation} and Theorem \ref{theorem:separation-FP}, we
obtain the following corollary. This shows a quite different nature of
adaptive and nonadaptive queries.

\begin{corollary}
There are two oracles $A$ and $B$ such that
$\eqp_{\|}^{A}\nsubseteq\p^{A}$ and $\p^{B}\nsubseteq\eqp_{\|}^B$.
\end{corollary}

\section{Applications to Decision Problems}

The study of decision problems has been extensively conducted in
quantum complexity theory and has brought in fruitful results
\cite{BBBV97,BV97,FGHP99,FR99,Wat00,YY99}. These results address the 
strengths and weaknesses of quantum computations. For instance, $\nqp$
characterizes $\co\cequalp$ \cite{FR99,FGHP99,YY99}, $\bqp$ is
contained within $\mathrm{AWPP}$ \cite{FR99}, and any $\pspace$-set
has a polynomial-time quantum interactive proof system
\cite{Wat99,KW00}. This section demonstrates two applications of
quantum functions to decision problems and makes a bridge between
language classes and function classes.

\subsection{A Quantum Characterization of PP}

Many quantum complexity classes lie within the probabilistic
complexity class $\pp$. This class $\pp$ is known to be robust since
it is characterized in many different fashions. For example, $\pp$ is
characterized by two $\gapp$-functions; namely, $\pp$ equals the
collection of all sets $A$ such that there exist two $\gapp$-functions
$f$ and $g$ satisfying that, for every $x$, $x\in A$ iff
$f(x)>g(x)$. We use a series of results in the previous sections to
show a new characterization of $\pp$ in terms of
$\sharpqp_{\algebraic}$ and $\gapqp_{\algebraic}$.

\begin{theorem}\label{theorem:PP}
Let $A$ be any subset of
$\Sigma^*$. The following statements are all equivalent.
\vs{-2}
\begin{enumerate}
\item $A$ is in $\pp$.
\vs{-2}
\item There exist two functions $f,g\in\sharpqp_{\algebraic}$ such that, 
for every $x$, $x\in A$ iff $f(x)>g(x)$.
\vs{-2}
\item There exist two functions $f,g\in\gapqp_{\algebraic}$ such that, 
for every $x$, $x\in A$ iff $f(x)>g(x)$.
\end{enumerate}
\end{theorem}

\begin{proof}
1 implies 3) Since $A\in \pp$, there exist a polynomial-time
deterministic TM $M$ and a polynomial $p$ such that, for every $x$,
$x\in A$ iff $|\{y\in\Sigma^{p(|x|)}\mid M(x,y)=1\}|>
2^{p(|x|)-1}$. Let $h(x)= |\{y\in\Sigma^{p(|x|)}\mid M(x,y)=1\}|$ for
every $x$. By modifying the proof of Lemma \ref{lemma:inclusion}, we
can show the existence of a unique function
$f\in\sharpqp_{\algebraic}$ satisfying that $h(x)=f(x)2^{p(|x|)}$ for
every $x$. Therefore, $x\in A$ iff $f(x)>\frac{1}{2}$. Define
$g(x)=\frac{1}{2}$ for all $x$. Clearly, $g$ is in
$\sharpqp_{\algebraic}$. Since
$\sharpqp_{\algebraic}\subseteq\gapqp_{\algebraic}$, claim 3) follows.

3 implies 2) Assume that there exist two
$\gapqp_{\algebraic}$-functions $f$ and $g$ such that $A=\{x \mid
f(x)>g(x)\}$. Using Proposition \ref{prop:gap-character}, take four
functions $k_0,k_1,h_0,h_1\in\sharpqp_{\algebraic}$ satisfying that
$f=k_0-h_0$ and $g=k_1-h_1$. Define
$\tilde{f}(x)=\frac{1}{2}(k_0(x)+h_1(x))$ and $\tilde{g}(x)=
\frac{1}{2}(k_1(x)+h_0(x))$ for all $x$. Lemma
\ref{lemma:operators}(2) guarantees that $\tilde{f}$ and $\tilde{g}$ 
are in $\sharpqp_{\algebraic}$. It is also obvious that $f(x)>g(x)$
iff $\tilde{f}(x)>\tilde{g}(x)$. Thus, we have $A=\{x\mid
\tilde{f}(x)>\tilde{g}(x)\}$.

2 implies 1) Assume that there exist two functions $f$ and $g$ in
$\sharpqp_{\algebraic}$ such that $A=\{x\mid f(x)>g(x)\}$. Define
$h(x)=f(x)-g(x)$ for all $x$. It follows from Proposition
\ref{prop:gap-character} that $h$
belongs to $\gapqp_{\algebraic}$. Moreover, by Theorem
\ref{theorem:property}(3), there exists a function $k$ in $\gapp$ such
that $sign(h(x))= sign(k(x))$ for all $x$. This implies that $x\in A$
iff $k(x)>0$. {}From the $\gapp$-characterization of $\pp$, it follows
that $A$ is in $\pp$.  
\end{proof}

To see the robustness of $\pp$, we consider the quantum analogue of
$\pp$.

\begin{definition}\label{def:PQP}
Let $\pqp_{K}$ be the collection of all sets $A$ such that there
exists a polynomial-time well-formed QTM with $K$-amplitudes
satisfying: for every $x$, if $x\in A$ then $M$ accepts $x$ with
probability more than $1/2$, and if $x\not\in A$ then $M$ accepts $x$
with probability at most $1/2$.
\end{definition}

{}From the above definition, we immediately obtain that
$\bqp_{K}\subseteq\pqp_{K}$. Thus, $\pqp_{\complex}$ has uncountable
cardinality since so does $\bqp_{\complex}$ \cite{ADH97}. This
concludes that $\pqp_{\complex}\neq\pp$. In contrast, any
$\pqp_{K}$-set $A$ has the form $A=\{x\mid f(x)>0\}$ for a certain
$\gapqp_{K}$-function $f$. Theorem
\ref{theorem:PP} then implies that, when $K$ is limited to $\algebraic$,
this $A$ falls into $\pp$. Overall, we obtain the following.

\begin{proposition}\label{prop:PQP-PP}
$\pqp_{\algebraic} = \pp$ and $\pqp_{\complex}\neq\pp$.
\end{proposition}

For the amplitude set $\ptcomplex$, Theorem \ref{theorem:property} is
not sufficient to conclude that $\pqp_{\ptcomplex}=\pp$. It is unknown
even whether $\pqp_{\ptcomplex}$ equals $\co\pqp_{\ptcomplex}$. It
seems, however, difficult to show the separation between
$\pqp_{\ptcomplex}$ and $\co\pqp_{\ptcomplex}$ since this immediately
implies the unproven consequence $\eqp_{\ptcomplex}\neq \cequalp$.

\begin{lemma}
$\pqp_{\ptcomplex}\neq\co\pqp_{\ptcomplex}$ implies
$\eqp_{\ptcomplex}\neq \cequalp$.
\end{lemma}

\begin{proof}
We show the contrapositive. We omit script $\ptcomplex$ for
readability. Assume that $\eqp=\cequalp$. Let $A\in\pqp$. There exists
a function $f\in\gapqp$ satisfying that $A=\{x\mid f(x)>0\}$. The
Squared Function Theorem implies that $f^2\in\sharpqp$. Consider the
function $g$ defined by $g(x)=1$ if $f^2(x)=0$ and $g(x)=-f(x)$
otherwise. Let $B=\{x\mid f^2(x)=0\}$. By the
$\sharpqp$-characterization of $\nqp$, $\overline{B}$ belongs to
$\nqp$ and thus, $B$ is in $\cequalp$. It is easy to show that $g$ is
in $\gapqp^{B[1]}$ by making a single query ``$x\in?B$'' and then
computing $f(x)$ (if necessary). By our assumption, $g\in
\gapqp^{\cequalp[1]}\subseteq\gapqp^{\eqp}$, which is $\gapqp$ by
Lemma \ref{lemma:query-FEQP}(5). Note that, for every $x$, $x\in A$
implies $g(x)<0$ and $x\not\in A$ implies $g(x)>0$. This concludes
that $A$ is in $\co\pqp$. Therefore,
$\pqp\subseteq\co\pqp$. Symmetrically, we can show that
$\co\pqp\subseteq\pqp$.  
\end{proof}

\subsection{Closure Properties of \#QP}

The closure properties of $\sharpp$ under various polynomial-time
computable operators were studied in \cite{OH93}. Such closure
properties imply the collapse of certain complexity classes, such as
$\up$ and $\spp$. Let $\circ$ be any operator between two functions. A
function class $\FF$ is said to be {\em closed under operator $\circ$}
if, for every pair $f,g\in\FF$, $f\circ g$ is also in $\FF$. The
maximum operator $\max$ is defined by $\max\{f,g\} = \lambda
x.\max\{f(x),g(x)\}$ and the minimum operator $\min$ is defined by
$\min\{f,g\} =\lambda x.\min\{f(x),g(x)\}$. Ogihara and Hemachandra
\cite{OH93} showed that if $\sharpp$ is closed under the minimum
operator then $\np=\up$. This consequence can be changed to
$\cequalp=\spp$ if we assume that $\sharpp$ is closed under either the
minimum operator or the maximum operator \cite{OH93}.

We consider the closure property of $\sharpqp_{K}$ under the maximum
and minimum operators. In connection to this closure property, we
first introduce a new complexity class.  Hereafter, we identify a
binary string with a rational number (expressed as a pair of two
integers but not as a dyadic number): for example,
$f(x)=\frac{2}{3(|x|+1)}$.

\begin{definition}\label{def:WQP}
A set $A$ is in $\wqp_K$ {\em(wide QP)} if there exist two functions
$f\in\sharpqp_K$ and $g\in\feqp_K$ with
$\ran(g)\subseteq(0,1]\cap\rational$ satisfying that
$f(x)=\chi_{A}(x)\cdot g(x)$ for every $x$.
\end{definition}

Notice that we can replace $\sharpqp_K$ in Definition \ref{def:WQP} by
$\gapqp_K$ if $K$ is admissible. Moreover,
$\eqp_K\subseteq\wqp_K\subseteq\nqp_K$ for any amplitude set $K$.
Now, we show the following proposition.

\begin{proposition}
Let $K$ be any admissible set.
\vs{-2}
\begin{enumerate}
\item If $\eqp_{K}=\pqp_{K}$, then $\sharpqp_{K}$ is closed under 
the maximum and minimum operators.
\vs{-2}
\item If $\sharpqp_{\algebraic}$ is closed under the maximum and 
minimum operators, then $\wqp_{\algebraic}=\pp$.
\end{enumerate}
\end{proposition}

\begin{proof}
1) Assume that $\eqp_{K}=\pqp_{K}$. Let $g$ and $h$ be any two
functions in $\sharpqp_{K}$ and set $f=\max\{g,h\}$. Define $A=\{x\mid
g(x)> h(x)\}$. By Proposition \ref{prop:gap-character}, the function
$\tilde{g}$ defined by $\tilde{g}=g-h$ is in $\gapqp_{K}$. Since
$A=\{x\mid \tilde{g}(x)>0\}$, $A$ belongs to $\pqp_{K}$. By our
assumption, $A$ is also in $\eqp_{K}$.  It is obvious that $f$ belongs
to $\sharpqp_{K}^{A[1]}$, which is a subset of
$\sharpqp_{K}^{\eqp_{K}}$. Since $K$ is admissible, we can show that
$\sharpqp_{K}^{\eqp_{K}} =\sharpqp_{K}$ similar to Lemma
\ref{lemma:query-FEQP}(4). Hence, $f$ is in $\sharpqp_{K}$. This
implies that $\sharpqp_{K}$ is closed under $\max$.  Similarly, we can
show the case for the minimality.

2) Assume that $\sharpqp_{\algebraic}$ is closed under $\max$ and
$\min$. Let $A$ be any set in $\pp$. By Proposition \ref{prop:PQP-PP},
$A$ belongs to $\pqp_{\algebraic}$ and thus, there exists a quantum
function $f\in\sharpqp_{\algebraic}$ such that $A=\{x\mid
f(x)>1/2\}$. Let $h=\max\{f,\frac{1}{2}\}$. By our assumption, $h$ is
in $\sharpqp_{\algebraic}$. Note that $\lambda x.(h(x)-\frac{1}{2})$
is in $\gapqp_{\algebraic}$. Now, define $k(x)=(h(x)-\frac{1}{2})^2$
for all $x$. By the Squared Function Theorem, $k$ is in
$\sharpqp_{\algebraic}$. Since $k\in\sharpqp_{\algebraic}$, take an
appropriate polynomial $p$ such that $k(x)\geq 2^{-p(|x|)}$ for all
$x$ (this fact is implicitly used in the proof of Theorem
\ref{theorem:property}(3)). Finally, we define $j=\min\{k,\lambda
x.2^{-p(|x|)}\}$. Since $\lambda x.2^{-p(|x|)}$ is in
$\sharpqp_{\algebraic}$, $j$ also belongs to $\sharpqp_{\algebraic}$
by our closure assumption of $\sharpqp_{\algebraic}$. This $j$
satisfies that $j(x)=\chi_{A}(x)\cdot 2^{-p(|x|)}$ for every $x$.
Thus, $A$ belongs to $\wqp_{\algebraic}$.  
\end{proof}

As is shown below, $\wqp$ is in some sense a generalization of $\up$.

\begin{lemma}\label{lemma:UP-WQP}
$\up\subseteq \wqp_{K}$ if $K\supseteq\{0,\pm1,\pm\frac{1}{2}\}$.
\end{lemma}

\begin{proof}
Take any set $A$ in $\up$. Note that $\chi_{A}\in\sharpp$. Lemma
\ref{lemma:inclusion} guarantees the existence of two functions
$f\in\sharpqp_{K}$ and $\ell\in\fp\cap\nat^{\Sigma^*}$ satisfying
$\chi_{A}(x)=\ell(1^{|x|})f(x)$ for all $x$. Define $g$ as follows:
for every $x$, $g(x)=\frac{1}{\ell(1^{|x|})}$ if $\ell(1^{|x|})\neq 0$
and $g(x)=1$ otherwise.  Thus, $f(x)=\chi_{A}(x) g(x)$ for all
$x$. Clearly, $\ran(g)\subseteq (0,1]\cap\rational$. Since
$g\in\fp\subseteq\feqp_{K}$, $A$ belongs to $\wqp_{K}$.  
\end{proof}

There exists a relativized world where $\eqp$ and $\wqp$ are different
classes. A {\em relativized WQP} is naturally introduced 
by the use of a relativized $\feqp$ and a relativized $\sharpqp$.

\begin{proposition}\label{prop:EQP-WQP}
There exists an oracle $A$ such that $\eqp^A\neq\wqp^A$. 
\end{proposition}

\begin{proof}
Note that $\up^A\subseteq \wqp^A$ for any oracle $A$ because the proof
of Lemma \ref{lemma:UP-WQP} relativizes. It suffices to show that
$\up^A\nsubseteq\eqp^A$ for a certain oracle $A$. This immediately
follows from the result of Fortnow and Rogers \cite{FR99}, who proved
that $\p^A=\bqp^A\neq\up^A\cap\co\up^A$ for a certain oracle $A$.
\end{proof}

\bs
\paragraph{Acknowledgments.}
The author is grateful to Andrew Yao and Yaoyun Shi for a stimulating
discussion on quantum computations at Princeton University.  He also
thanks Harumichi Nishimura for careful proof-checking and Marina
Sokolova for her kind assistance 
in the revision process of the conference
version of this paper.

\bibliographystyle{alpha}

\end{document}